\pdfoutput=1

\documentclass[letterpaper,twocolumn,10pt]{article}

\usepackage{usenix,epsfig}

\usepackage{amsmath}
\usepackage{graphicx}
\usepackage{booktabs}
\usepackage{amssymb}
\usepackage{latexsym}	
\usepackage{array}		
\usepackage{multirow}
\usepackage{subfigure}
\usepackage{algorithm}
\usepackage{algpseudocode}
\usepackage{threeparttable}
\usepackage{paralist}   
\usepackage{xspace}
\usepackage{color}
\usepackage{adjustbox}
\usepackage{balance}
\usepackage{wasysym}
\usepackage{flushend}  
\usepackage{authblk}
\usepackage{framed}

\usepackage[hyphens]{url}
\usepackage[pdfborder={0 0 0}, citecolor=blue, linkcolor=blue, urlcolor=black, colorlinks=true]{hyperref}

\newif\ifACM
\ACMtrue  

\newif\ifUSENIX
\USENIXtrue

\ifACM
\newcommand{\myfig}{Figure\xspace}
\else
\newcommand{\myfig}{Fig.\xspace}
\fi

\ifACM
\newcommand{\mysec}{\S}
\else
\newcommand{\mysec}{Section\xspace}
\fi

\newcommand{\red}[1]{\textcolor[rgb]{1.00,0.00,0.00}{{#1}}\xspace}
\newcommand{\blue}[1]{\textcolor[rgb]{0.00,0.00,1.00}{{#1}}\xspace}

\newcommand\comment[1]{}

\newcommand{\name}{MopEye\xspace}

\begin{document}

\date{}

\title{\Large \bf \name: Opportunistic Monitoring of Per-app Mobile Network Performance}

\author[$1$\thanks{Half of the work by this author was performed at The Hong Kong Polytechnic University.}]{\rm Daoyuan Wu}
\author[$2$]{\rm Rocky K. C. Chang}
\author[$2$]{\rm Weichao Li}
\author[$2$]{\rm Eric K. T. Cheng}
\author[$1$]{\rm Debin Gao}
\affil[$1$]{Singapore Management University}
\affil[$2$]{The Hong Kong Polytechnic University}
\affil[$$]{\red{This paper has been accepted by 2017 USENIX Annual Technical Conference (ATC'17).}}
\affil[$$]{\large\url{https://mopeye.github.io}\thanks{We thank Dr. Ada Gavrilovska for shepherding our paper and the anonymous reviewers for their valuable comments. This work is partially supported by a grant (ref. no. G-YBAK) from The Hong Kong Polytechnic University, a grant (ref. no. H-ZL17) from the Joint Universities Computer Centre of Hong Kong, and the Singapore National Research Foundation under NCR Award Number NRF2014NCR-NCR001-012.}}

\maketitle


\begin{abstract}

Crowdsourcing mobile user's network performance has become an effective way of understanding and improving mobile network performance and user quality-of-experience.
However, the current measurement method is still based on the landline measurement paradigm in which a measurement app measures the path to fixed (measurement or web) servers.
In this work, we introduce a new paradigm of measuring \textit{per-app} mobile network performance. We design and implement \name, an Android app to measure network round-trip delay for each app whenever there is app traffic. This opportunistic measurement
can be conducted automatically without user intervention. Therefore, it can facilitate a large-scale and long-term crowdsourcing of mobile network performance.
In the course of implementing \name, we have overcome a suite of challenges to make the continuous latency monitoring lightweight and accurate.
We have deployed \name to Google Play for an IRB-approved crowdsourcing study in a period of ten months, which obtains over five million measurements from 6,266 Android apps on 2,351 smartphones.
The analysis reveals a number of new findings on the per-app network performance and mobile DNS performance.

\end{abstract}

\vspace{-2ex}
\section{Introduction}
\label{sec:intro}
\vspace{-1ex}

In recent years, a number of
crowdsourcing platforms using smartphone
apps are deployed to measure mobile network performance. MobiPerf~\cite{mobiperf_android} and
Netalyzr~\cite{netalyzr_android} on Android, for example, enable users
to measure a number of network performance metrics between their
smartphones and remote endpoints. Using these uncoordinated
network measurement performed by end users to obtain accurate and
meaningful insights is still under active
research~\cite{Mobilyzer15}. Related to that, a number of speedtest
services are provided for Android~\cite{speedchecker_android,
  speedtest_android}, iOS~\cite{speedtestx_ios, speedtest_ios}, and
Windows Phone users~\cite{nspeedtest_win, speedtest_win}.

The existing mobile measurement apps, however, are still based on the
landline measurement paradigm. They actively send probe
packets to user-specified remote endpoints or measurement servers
(e.g., M-Lab servers). Due to the diverse locations of various servers and user
mobility, such landline measurement will not correlate well with the
user's experience.
In this paper, we propose to measure mobile network performance for each app (i.e., from user's smartphone to the app server). The \textit{per-app} measurement not only reflects user's experience with the app but also helps diagnose application-specific problems.
An effective approach to per-app measurement is to perform the measurement only when there is app traffic. Since this opportunistic measurement can be conducted automatically without user's intervention, it can facilitate a large-scale and long-term crowdsourcing of mobile network performance.

\begin{figure}[t!]
\vspace{-2ex}
\begin{adjustbox}{center}   
  \subfigure[\small An all-app view.] {
	\label{fig:mainUI}
    \includegraphics[width=0.26\textwidth]{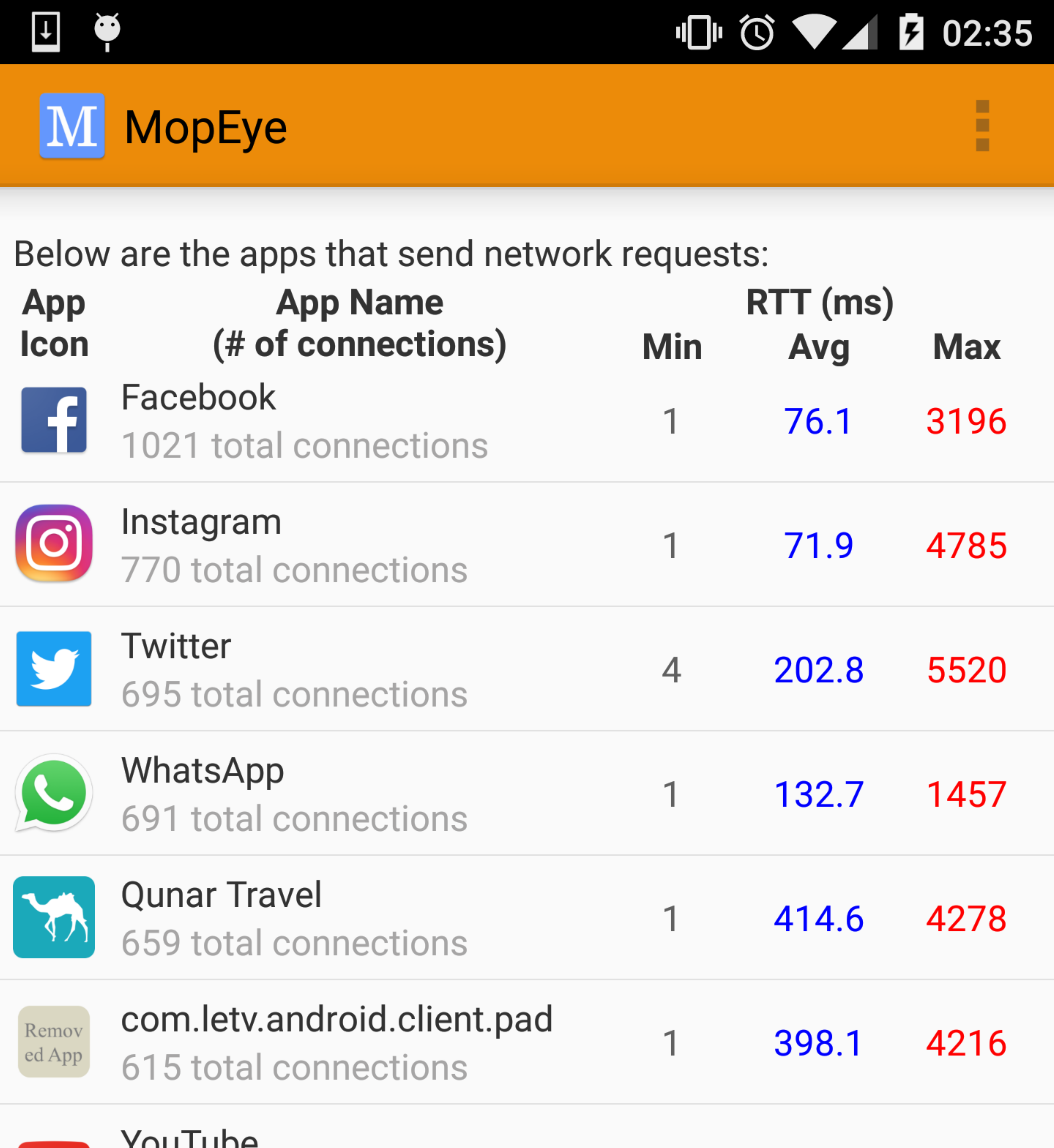}
  }
  \hspace{-2ex}            
  \subfigure[\small An individual-app view.] {
	\label{fig:secondUI}
    \includegraphics[width=0.26\textwidth]{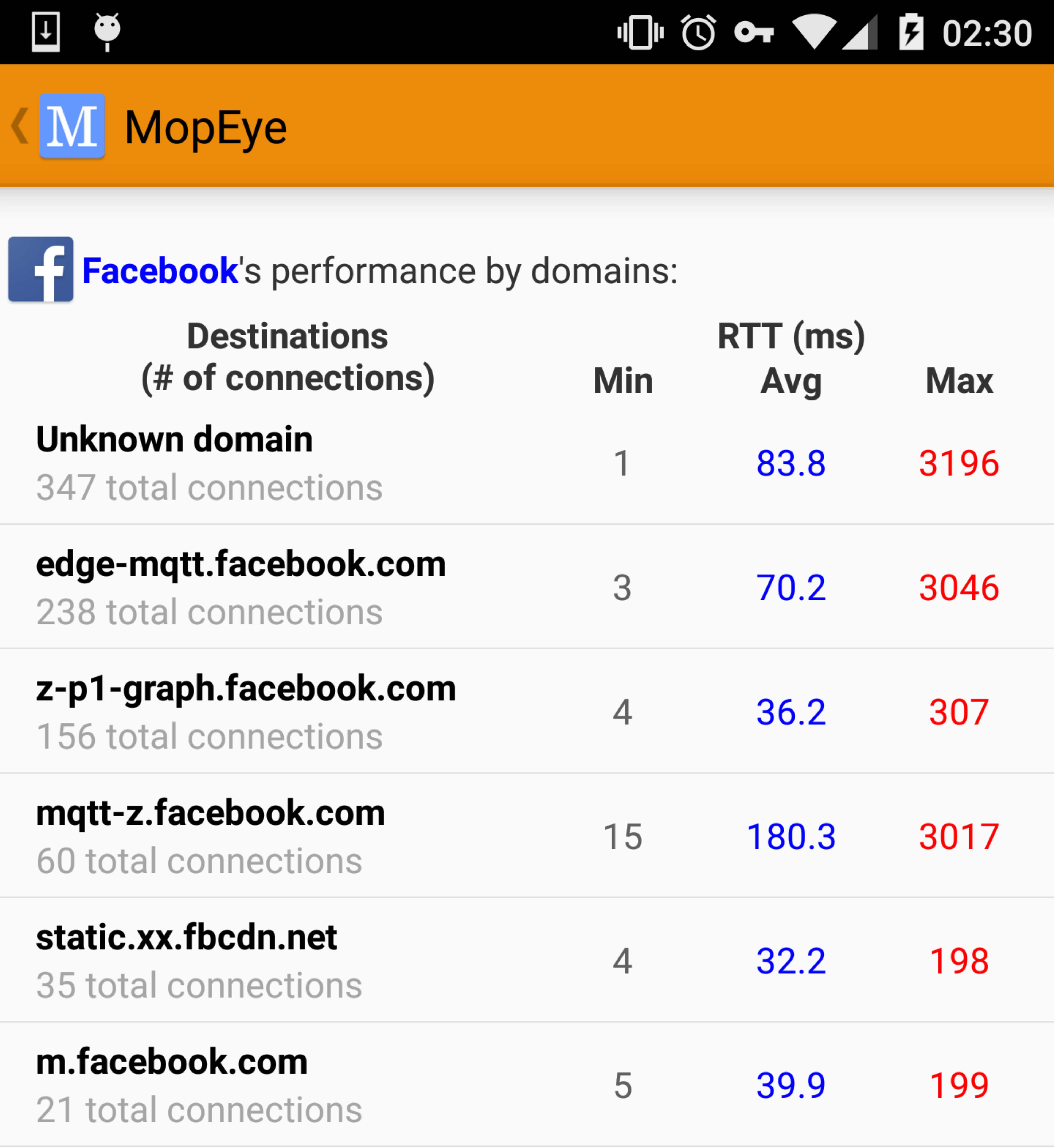}
  }
\end{adjustbox}
\vspace{-3ex}
\caption{\name's two major user interfaces.}
\vspace{-2ex}
\label{fig:toolUI}
\end{figure}

In this paper, we utilize the \texttt{VpnService} API
available on Android 4.0+~\cite{AndroidVPNAPI} to implement opportunistic measurement of per-app network performance
in \name (MObile Performance Eye), our Android measurement app.
\myfig~\ref{fig:toolUI} shows the two main interfaces in \name.
With the \texttt{VpnService} interface, \name can \textit{passively} capture the traffic initiated
by all apps and forward them \textit{actively} to the remote app servers
using socket calls. Based on the \texttt{connect()} socket calls, it can
estimate the round-trip time (RTT) for each app. Therefore,
the measurement incurs zero network overhead, and the RTT can accurately
reflect the network delay experienced by each app. Moreover, \name can be deployed easily, because it
does not need the root privilege which is required for \texttt{tcpdump}-based passive measurement.
It is also very easy to operate. Users are only required to consent to enabling \name's VPN interface once. After that, MopEye performs the measurement opportunistically and autonomously.

The main challenge in the design and implementation of \name is to
mitigate the impact on other apps by performing \textit{fast} packet
relaying. However, our design choices are constrained by two important restrictions: no
relaying using a remote VPN server and no raw sockets which require
the root privilege. To satisfy the constraints, we build our own user-space TCP/IP stack to
perform packet relaying between the VPN tunnel packets and
those in the socket connections. In particular, we have identified and overcome a number of serious performance
degradation issues in the entire packet-relaying process.
Another challenge is to obtain high measurement accuracy. Based on our evaluation, MopEye's mean RTT
measurement deviates from \texttt{tcpdump}'s results by at most 1ms.
Besides that, our evaluation also shows that \name incurs very low overhead on the throughput, battery consumption, and CPU usage.

We have deployed \name to Google Play~\cite{MopEyeOnPlay} for an IRB-approved crowdsourcing study since May 2016.
We have so far\footnote{By the time of our submission on 7 February 2017.} attracted 4,014 user installs from 126 countries and collected the \textit{first} large-scale per-app measurement dataset comprising 5,252,758 RTT measurements from 6,266 Android apps on 2,351 smartphones\footnote{Note that many users use daily apps such as Facebook and Whatsapp. Thus, there is a large common app space among different phones.}.
An analysis of these crowdsourced data reveals a number of new findings on the per-app and DNS network performance experienced by real users under different network types and ISPs in the wild.
We also perform several case studies to diagnose the performance issues in Whatsapp, India's largest 4G ISP, and two American cellular ISPs.

\vspace{-2ex}
\section{Design of MopEye}
\label{sec:design}
\vspace{-2ex}

In this section, we present an overview of \name and its main
components. We defer the implementation details and performance enhancement to the
next section.

\subsection{\name Overview}
\label{sec:overview}

\begin{figure}[t!]
\vspace{-2ex}
\begin{adjustbox}{center}
\includegraphics[width=0.48\textwidth]{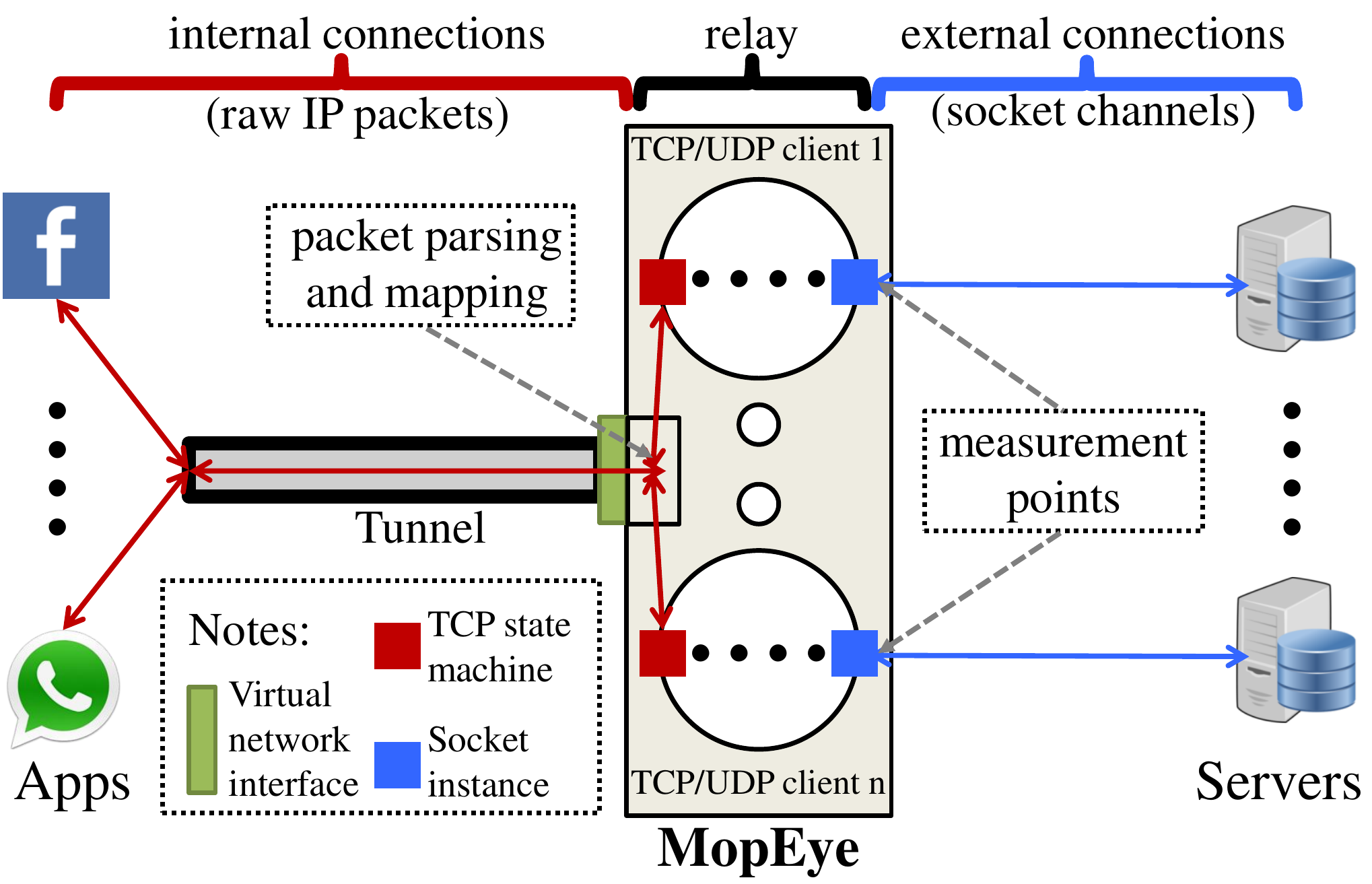}
\end{adjustbox}
\vspace{-4ex}
\caption{An overview of \name.}
\vspace{-2.3ex}
\label{fig:overview}
\end{figure}

\myfig~\ref{fig:overview} presents a high-level design of
\name. There are three main steps for \name to use an app's traffic to opportunistically
measure the network RTT. For the outgoing traffic, \name first
captures an app's packets through a tunnel, relays the
captured packet to an external TCP connection or UDP association with a remote server, and sends
the packets to the server.
In the last step, \name calculates the time between the app's SYN and SYN/ACK packets to measure the RTT. 
The RTT measurement for UDP apps is similar (i.e., between query and response messages).
In the following we describe each step in more details.


\vspace{-2ex}
\subsection{Packet Capturing, Parsing, and Mapping}
\label{sec:mapping}
\vspace{-1ex}

We leverage Android's \texttt{VpnService} APIs to build a virtual
network interface (green box in \myfig~\ref{fig:overview}) to
intercept all traffic initiated from any app on the smartphone. It
also receives server-initiated traffic, but for the sake of simplicity we do not
discuss this traffic direction in this paper.

\begin{figure}[t!]
\vspace{-2ex}
\begin{adjustbox}{center}
\includegraphics[width=0.35\textwidth]{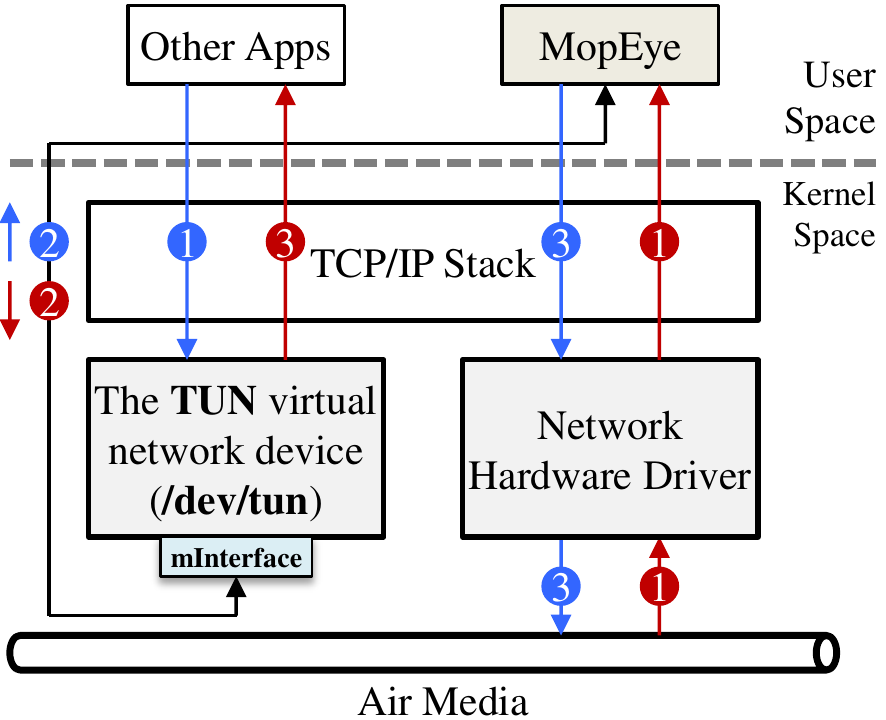}
\end{adjustbox}
\vspace{-5ex}
\caption{\small \name's packet capturing and relaying for incoming flow (red) and outgoing flow (blue). The black link represents a bi-directional flow.}
\vspace{-3.8ex}
\label{fig:mechanism}
\end{figure}

Android's \texttt{VpnService} APIs leverage the \texttt{TUN} virtual
network device (\texttt{/dev/tun} on Android or \texttt{/dev/net/tun}
on some UNIX systems) to capture packets. \myfig~\ref{fig:mechanism}
illustrates \name's packet capturing and relaying mechanisms for the
incoming and outgoing traffic. Once \name builds a \texttt{TUN}
interface (i.e., \texttt{mInterface} in the figure), the \texttt{TUN}
device driver will capture and deliver all outgoing app packets to
this interface. \name then obtains these packets using
\texttt{mInterface}'s input stream. It is worth noting that the
packets captured here are all IP packets, because a \texttt{TUN}
device is essentially a virtual point-to-point IP link. \name
parses the captured packets to obtain the IP addresses and port
numbers for packet relaying.

To support per-app measurement, \name must also determine to which app
a captured packet belongs. Although there is no API support for this
socket-to-app mapping function, we find that four pseudo
files in the \texttt{proc} filesystem
(\texttt{/proc/net/tcp6|tcp|udp|udp6}) store each TCP/UDP connection's
local and remote IP addresses and ports, as well as the corresponding
app's UID which is a unique ID for each installed app. Moreover, using
Android's \texttt{PackageManager} APIs, \name obtains the app's name
from its UID. To reduce the overhead of this procedure, \name performs
this operation only for the SYN packets, and the resolved names and
socket addresses are cached for the subsequent data
packets. Furthermore, we will present in \mysec\ref{sec:lazymap} a
new mechanism to significantly minimize the mapping overhead even for
SYN packets.
As for UDP packets, \name currently supports only DNS measurement (though it relays all UDP packets).
Since DNS is system-wide, \name does not need to map UDP packets for now.

\vspace{-2ex}
\subsection{Packet Relaying}
\label{sec:relaying}
\vspace{-1ex}

Relaying packets between apps and their servers efficiently is the
most challenging task in the design and implementation of \name. Our
solution to this problem is shaped by the three main considerations
below.
\begin{compactitem}
\item \textbf{Measurement objective} Since our goal is to measure the
  RTT between a user's smartphone and the app servers, we cannot rely
  on a remote VPN server to relay the application packets to their
  servers. Therefore, we require \name to relay packets within the
  smartphone.

\item \textbf{Running on unrooted phones} Our another objective is to
  run \name on unrooted phones. Using raw sockets to relay packets to
  the servers is therefore not an option. Instead, \name must relay
  packets via the regular TCP/UDP sockets for the external
  connections. We have implemented both TCP and UDP packet relays. Due
  to the page limit, we describe only the TCP relay from now on.

\item \textbf{User-space TCP stack} As a result of using regular TCP
  socket, \name will not be able to access the information in the TCB
  (Transmission Control Block~\cite{RFC793}), such as the TCP sequence
  and acknowledgement numbers, from the external
  connections. Therefore, \name must create its own user-space TCP
  stack (in the form of TCP state machine) for the internal
  connections. We refer the packets transmitted in the internal and
  external connections to as \textit{tunnel packets} and \textit{socket
    packets}, respectively.
\end{compactitem}

\noindent\textbf{Splicing the two connections} To relay packets in a
TCP connection, \name ``splices'' the internal connection terminated
by \name's TCP state machine and the external connection initiated by
\name's TCP socket. Our approach is to link the state machine and the
socket with two-way referencing. That is, we create a TCP client
object that wraps the socket instance and include a reference to the
state machine. The state machine also maintains a reference to the
corresponding TCP client.

\noindent\textbf{Processing tunnel packets}
\name processes the tunnel packets according to RFC
793~\cite{RFC793}. The processing logics for different TCP
packets are summarized as follows.
\begin{compactitem}

\item \textit{TCP SYN:} Upon receiving a SYN packet, \name creates a
  TCP client object and uses its socket instance to perform handshake
  with the remote server. Only after establishing the external
  connection can \name complete the handshake with the app.

\item \textit{TCP Data:} \name places the data from tunnel packets to
  a socket write buffer and triggers a socket write event for the
  socket instance to handle.

\item \textit{Pure ACK:} \name discards pure ACK packets, because
  there is no need to relay them to the socket channel.

\item \textit{TCP FIN:} \name updates the TCP state to half closed and
  generates an ACK packet to the app. Meanwhile, it triggers a
  half-close write event for the socket instance to handle.

\item \textit{TCP RST:} \name closes the external socket connection
  and removes the corresponding TCP client object from the cached TCP
  client list.

%
%

\end{compactitem}

\noindent\textbf{Processing socket packets} To handle concurrent
socket instances, \name uses \textit{non-blocking}
\texttt{SocketChannel} APIs to communicate with the remote app servers. In
particular, it uses a socket selector~\cite{JavaSelector} to listen
for read and write events, and handles them as follows.
\begin{compactitem}

\item \textit{Socket Read:} Upon detecting a read event, \name
  retrieves the incoming data from the read buffer and constructs data
  packets for the internal connection. In \mysec\ref{sec:tunetcp}, we
  propose a method to improve the performance of this step. However,
  if this read event is for a socket close/reset, \name generates a
  FIN/RESET packet for the internal connection.

\item \textit{Socket Write:} Upon detecting a write event, the socket
  instance sends all the data in the write buffer to the remote server
  and instructs the corresponding TCP state machine to generate an ACK
  packet to the app. However, if this write event is for half-close,
  \name closes the external connection and generates a FIN
  packet to the app.


\end{compactitem}

\subsection{Measurement Methods}
\label{sec:measuremethod}
\vspace{-1ex}

Obtaining accurate per-app RTT measurement using \name faces more
challenges than that using the traditional active measurement apps, such as
MobiPerf~\cite{mobiperf_android} and Ookla
Speedtest~\cite{speedtest_android}. There are two main challenges.
\begin{compactitem}

\item [\textit{C1:}] Since \name has no control on the relayed
  packets, it cannot execute pre-negotiated measurement logic as in
  active measurement apps. This challenge is further exacerbated due
  to the lack of TCB information for correlating packets for
  measurement.

\item [\textit{C2:}] Unlike other apps that have a relatively
  ``clean'' measurement environment, the performance and accuracy of
  \name can be easily affected by measurement noises, because it has
  to relay packets for all applications in the phone.

\end{compactitem}

To address challenge C1, we identify and correlate the correct packets for computing the RTT.
Among the four
types of TCP socket calls (i.e., \texttt{connect()}, \texttt{read()},
\texttt{write()}, and \texttt{close()}), our evaluation using
\texttt{tcpdump} shows that the \texttt{connect()} call always
accurately corresponds to a single round of packets, i.e., the SYN and SYN-ACK
pair. That is, invoking a \texttt{connect()} call will immediately
send out a SYN packet, and the call returns just after receiving a
SYN-ACK packet. In contrast, a \texttt{read()}/\texttt{write()} call
may involve multiple rounds of packet exchanges, and a
\texttt{close()} call may not always elicit an ACK packet from the
server.

However, it is difficult for \name to obtain the
post-\texttt{connect()} timestamp accurately due to
C2. Since \name uses non-blocking \texttt{SocketChannel} APIs to relay
packets, it has to wait for the system's notification for a received
ACK. This event-based notification can introduce an
additional delay up to several milliseconds if there are other pending
socket events (e.g., read/write or \texttt{VpnService}'s incoming
packets). We resolve this inaccuracy problem by temporarily setting the
socket into blocking mode for each \texttt{connect()} call. That is,
\name runs a \texttt{connect()} call in a temporary new thread, which
we call \texttt{socket-connect} thread. Once the connection is
established, \name resumes the non-blocking mode and switches back to
the main thread listening for read and write events. As a result,
\name can obtain an accurate post-\texttt{connect()} timestamp for the
RTT measurement and, at the same time, provides efficient packet
relaying. As will be explained in \mysec\ref{sec:implement}, the
temporary \texttt{socket-connect} threads also give us several other
benefits for optimizing \name's performance.

Besides the TCP-based measurement, \name also supports DNS.  Measuring
the RTT for DNS is quite straightforward. We can obtain it by
measuring the time between \texttt{send()} and \texttt{receive()} UDP
socket calls, which correspond to DNS query and reply, respectively.
However, obtaining an accurate post-\texttt{receive()} timestamp is
still difficult because of C2.  We adopt a similar solution
by setting up a temporary thread for a blocking-mode measurement,
except that this time we run the whole DNS processing, including DNS
parsing and socket initialization, in the temporary thread (instead of
just doing so for the \texttt{connect()} call as in the TCP
measurement).  This is because DNS is an application-layer protocol
built upon UDP, and processing it should not block the main thread of
\texttt{VpnService}.

\vspace{-2ex}
\section{Implementation and Enhancements}
\label{sec:implement}
\vspace{-1ex}

\begin{figure*}[t!]
\vspace{-4ex}
\begin{adjustbox}{center}
\includegraphics[width=0.85\textwidth]{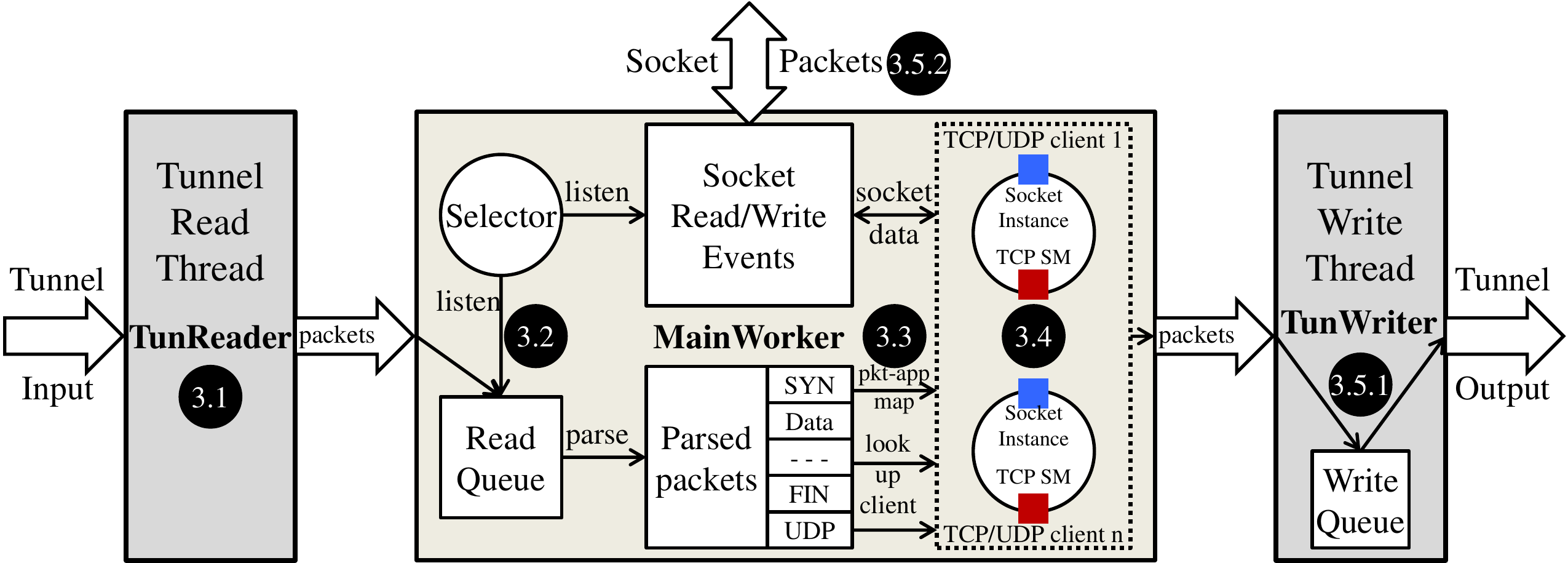}
\end{adjustbox}
\vspace{-4ex}
\caption{The architecture of \name.}
\vspace{-3ex}
\label{fig:impleoverview}
\end{figure*}

We have implemented \name in 11,786 LOC and deployed it to Google Play~\cite{MopEyeOnPlay} on 16 May 2016 for a crowdsourcing measurement study\footnote{IRB approval was obtained from Singapore Management University on 9 October 2015 under application IRB-15-093-A077(1015).}.
\myfig \ref{fig:impleoverview} presents the architecture of \name.  It
has three major components or core threads (created by our
\texttt{MopEyeService} that extends the Android \texttt{VpnService}
class). The \texttt{TunReader} and \texttt{TunWriter} threads handle
read/write for the VPN tunnel, whereas the \texttt{MainWorker} thread is
responsible for all the packet processing (i.e., packet parsing,
mapping, and relaying) and RTT measurement.

In this section, we will detail how we solve the challenges of
implementing \texttt{TunReader}, \texttt{TunWriter}, and
\texttt{MainWorker}, particularly our methods of enhancing \name's
performance.  For better reading and quick reference, we include the
subsection numbers in the corresponding components in
\myfig~\ref{fig:impleoverview}.
Among them, \mysec\ref{sec:tuninput} and \mysec\ref{sec:socketwrite} present solutions generic to all VPN-based apps on Android, whereas the rest can benefit various VPN-based traffic inspection systems on different OSes.

\vspace{-2ex}
\subsection{Zero-delay Packet Retrieval from the VPN Tunnel}
\label{sec:tuninput}
\vspace{-1ex}
Reading packets from the VPN tunnel is straightforward, but it is very
challenging to fast-retrieve the packets under the existing Android VPN
programming paradigm.  To illustrate this problem, we use a code snippet from
ToyVpn~\cite{ToyVpn}, a representative VPN
client in the official Android SDK sample code.
The code\footnote{Due to the page limit, we skip the code here and refer interested readers to \url{http://tinyurl.com/ToyVPN}.}
shows a 100ms sleep before executing each
\texttt{read()} call.  The purpose of this sleep is to reduce
CPU cycles for data reading. Therefore, the sleep period is
determined by the tradeoff between CPU consumption and packet
retrieval delay.

We are not aware of any solution addressing this delayed VPN read
problem. The ToyVpn example~\cite{ToyVpn} implements an
``intelligent'' sleeping algorithm to partially mitigate this
problem. The basic idea is to stop sleeping when detecting consecutive
packet reads. The recently proposed Haystack~\cite{Haystack15} adopts
a similar idea, but the system performance is not acceptable, e.g.,
achieving only 17.2Mbps throughput from a 73Mbps upload
link. PrivacyGuard~\cite{PrivacyGuard15}, another system using
\texttt{VpnService}, simply sets the sleep interval to 20ms.

We propose to fundamentally solve this problem by putting the VPN
\texttt{read()} API into a blocking mode.  That is, each
\texttt{in.read()} call will be blocked until a packet is retrieved
from the tunnel. This will effectively relieve the CPU from checking
for data continuously. As a result, we must run the VPN
\texttt{read()} API in a dedicated thread, i.e., \texttt{TunReader} in
\name, and the retrieved packets will be put in a read queue shown in
\myfig~\ref{fig:impleoverview}.

Unfortunately, there is no API provided for setting the blocking mode
of the VPN interface's file descriptor until Android 5.0.  To
implement our idea also for Android 4.0 to 4.4, we propose the following
two solutions.  First, at the native code level, we can invoke the
\texttt{fcntl()} API with the \texttt{F\_SETFL} command to set the
blocking mode.  Second, we can leverage Java reflection to invoke a
non-API function called \texttt{setBlocking} in the unexported
\texttt{libcore.io.IoUtils} class. We verify that this private
function exists on Android from its inception.

Although we can achieve zero-delay packet retrieval, there is a side
effect of not being able to \textit{timely} stop the
\texttt{TunReader} thread in a blocking mode.  We have tried the
\texttt{Thread.interrupt()} API, but it does not work because in the absence of incoming packets the \texttt{read()} call will be blocked.  To
address this issue, we send a dummy packet to the VPN tunnel to
release the blocked \texttt{read()} call.  The dummy packet can be
sent by \name itself for Android versions below 5.0.
For Android 5.0+, however, \name no longer
has the capability of letting its own packets go through the VPN
tunnel due to the need of calling \texttt{addDisallowedApplication(mopeye)}
to improve the performance (see \mysec\ref{sec:socketwrite}).
The only solution is to trigger a network request from other
apps.  After careful consideration, we use Android DownloadManager
APIs~\cite{DownloadManager} to stably trigger dummy download requests.

\vspace{-2ex}
\subsection{Monitoring Selector and Read Queue}
\vspace{-1ex}
As shown in \myfig~\ref{fig:impleoverview}, we use a socket selector
to listen for non-blocking read/write events from each socket instance
and a read queue for receiving tunnel packets from \texttt{TunReader}.
Being implemented as a single thread, \texttt{MainWorker}, however,
cannot monitor both the socket selector and the tunnel read queue at
the same time.
To circumvent this problem, we leverage the existing \texttt{select()}
waiting point to also monitor the read queue.  That is,
\texttt{TunReader} will issue a \texttt{Selector.wakeup()} event
whenever it adds a new packet to the read queue.
As a result, when the selector is woken up, \name will check for both
socket and tunnel events, because either could have activated the selector.
Moreover, to process the events timely, we interleave the code for
checking these two types of events.

\vspace{-2ex}
\subsection{Lazy Packet-to-App Mapping}
\label{sec:lazymap}
\begin{figure}[h!]
\vspace{-2ex}
  \begin{adjustbox}{center}
  \subfigure[Before the lazy mapping.] {
    \label{fig:beforelazy}
    \includegraphics[width=0.25\textwidth]{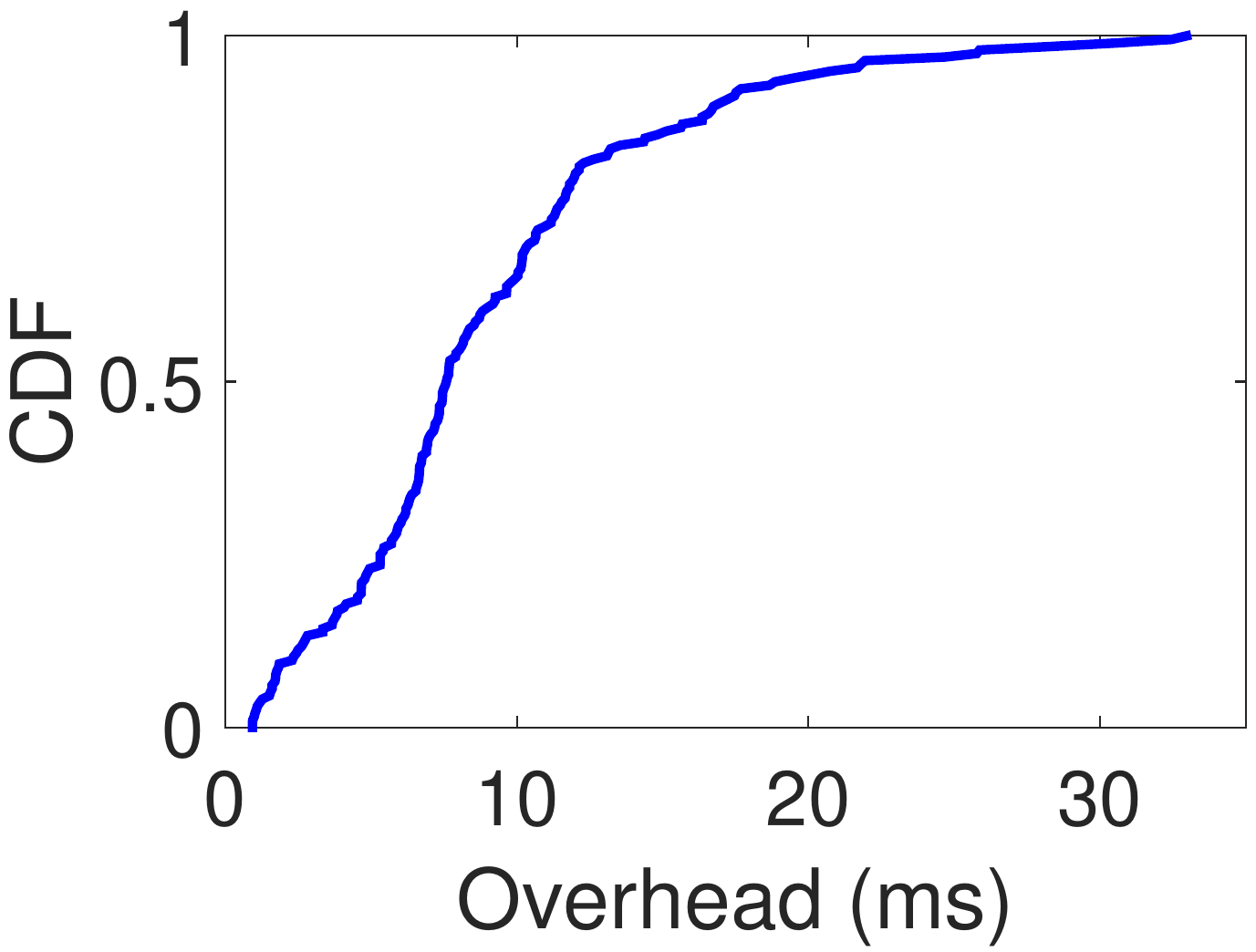}
  }
  \hspace{-2ex}
  \subfigure[After the lazy mapping.] {
	\label{fig:afterlazy}
    \includegraphics[width=0.25\textwidth]{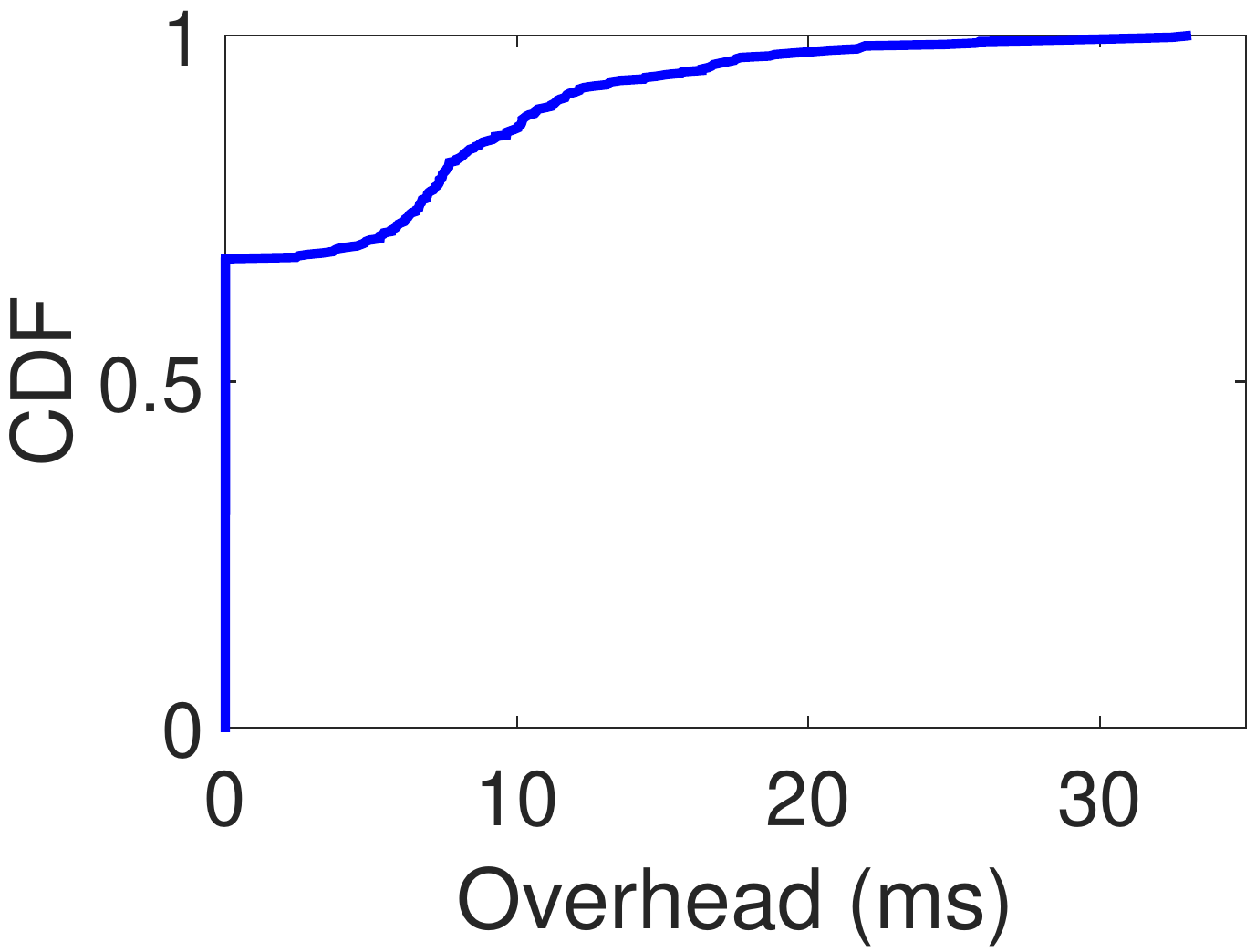}
  }
  \end{adjustbox}
\vspace{-4ex}
\caption{\small CDF plots of packet-to-app overhead per packet.}
\vspace{-1ex}
\label{fig:mapparseoverhead}
\end{figure}

\vspace{-1ex}
As presented in \mysec\ref{sec:mapping}, \name performs a
packet-to-app mapping for SYN packets in order to obtain per-app
network performance. Our evaluation, however, shows that such mapping
is expensive.  \myfig~\ref{fig:beforelazy} shows the cumulative
distribution function (CDF) of the overhead for parsing
\texttt{/proc/net/tcp6|tcp} for each SYN packet.  The experiment was
performed on a Nexus 6 phone, containing 196 samples, and in the experiment we browsed a list
of websites using the Chrome app. Over 75\% of the
samples required more than 5ms for the parsing; over 10\% of them needed
even more than 15ms. Furthermore, the overhead will increase with the number of
active connections in the system.

We propose a \textit{lazy} mapping mechanism to address this problem.
First, we defer the mapping from the main thread to each temporary
\texttt{socket-connect} thread mentioned in \mysec\ref{sec:measuremethod}.  Moreover, the mapping is performed only after the connection is established or failed, thus not
affecting the timely TCP handshake on the application side.  Second
and more importantly, we develop an efficient mapping algorithm that
performs less \texttt{proc} file parsing.  Our mapping algorithm is
based on the observation that for multiple concurrent
\texttt{socket-connect} threads, it is sufficient to let only one
thread perform the parsing. Other threads just check and/or
sleep to wait for the working thread to retrieve the mappings
for them. We choose the sleep period of 50ms which is sufficiently
large when compared with the parsing overhead in
\myfig~\ref{fig:beforelazy}.
The evaluation results show that such a \textit{lazy} mapping
algorithm is very useful for scenarios like web browsing.  For a total
of 481 temporary \texttt{socket-connect} threads in a web browsing
scenario, only 155 of them need to perform parsing.  Moreover, the
algorithm helps avoid the mapping overhead in the other 326 threads,
i.e., achieving 67.8\% mitigation rate as shown in \myfig \ref{fig:afterlazy}.
Besides improving the mapping
performance, it also helps reduce the CPU overhead.

Haystack~\cite{Haystack15} briefly mentions
that they use \textit{cache} to minimize the mapping overhead.
However, cache-based mechanism could cause
\textit{inaccurate} packet-to-app mapping results.  For example, both
the Facebook app and accessing Facebook by Chrome may use the same
server IP and port, but their mappings are different.  This problem is
more noticeable for advertisement modules since the same library may be
embedded in many different mobile apps.  Therefore, in order to obtain
an accurate mapping, we use our own lazy mapping mechanism instead of the
traditional cache-based mechanism.

\vspace{-1ex}
\subsection{Tuning TCP Performance}
\label{sec:tunetcp}
\vspace{-1ex}

Besides implementing the basic user-space TCP/IP stack presented in
\mysec\ref{sec:relaying}, we have identified and tuned the following
performance issues for fast packet relaying.

\noindent \textbf{Maximum segment size (MSS)} To maximize the
throughput of the internal connections, \name sets the MSS option to
1460 bytes in the SYN/ACK packet and sends 1500-byte IP packets to the
apps.

\noindent \textbf{Receive window size} Another factor affecting TCP
throughput is the TCP receive window. \name assigns the maximum of
65,535 bytes to each \name's socket write and read buffer. We could
also use the TCP window scale option~\cite{RFC1323} to further
increase the throughput but have not done so, because the existing
receive window is already big enough for achieving good performance and a
bigger window size will increase the buffer memory.

\noindent \textbf{No congestion and flow control} Since no packet loss
and reordering is expected in the VPN tunnel, \name forwards the data
packets continuously to the app without waiting for the
ACKs. Moreover, upon receiving a FIN/RST packet, \name stops the
packet forwarding immediately.

\noindent \textbf{Minimizing the use of expensive calls} We try to
minimize the use of expensive calls during the packet processing.  For
example, we discover that the \texttt{register()}
call~\cite{SocketRegister} for registering the socket selector can
sometimes be very expensive. \name therefore executes this call in the
\texttt{socket-connect} thread \textit{only after} completing the
handshaking for the internal connection. Other examples include never
performing database operations in the main thread and always avoiding
the debug log output.

\vspace{-2ex}
\subsection{Fast Dispatching of Tunnel and Socket Packets}
\label{sec:tunoutput}
\vspace{-1ex}

\subsubsection{Dispatching Packets to the VPN Tunnel}
\label{sec:tunnelwrite}
\vspace{-1ex}
We observe that writing packets to the tunnel is \textit{not} always
fast, partially because multiple writing threads (e.g.,
\texttt{MainWorker} and individual \texttt{socket-connect} thread) share only one tunnel.  We use the experimental results obtained
from two writing schemes in Table~\ref{tab:writeoverhead} to
illustrate this problem.
\begin{compactitem}
\item \textit{directWrite}: Writing is performed whenever there are
  packets to be sent to the tunnel.
\item \textit{queueWrite}: As illustrated in
  \myfig~\ref{fig:impleoverview}, the packets are first put in a
  queue. A separate writing thread is used to output the
  packets. This scheme is currently adopted by \name.
\end{compactitem}

\begin{table}[th!]
\vspace{-1ex}
\begin{adjustbox}{center}
\scalebox{
0.9
}{
\begin{tabular}{ |c || c | c || c | c| }

\hline
          & directWrite & queueWrite & oldPut   & newPut \tabularnewline
\hline
\hline

Total     & 1,244     & 2,161     & 810       & 5,321\tabularnewline
\hline
0$\sim$1ms  & 1,202     & 2,147     & 763       & 5,317 \tabularnewline
\hline
1$\sim$2ms  & 30        & 12        & 39        & 1\tabularnewline
\hline
2$\sim$5ms  & 7         & 2         & 7         & 1\tabularnewline
\hline
5$\sim$10ms & 3         & 0         & 1         & 2\tabularnewline
\hline
$>$10ms     & 2         & 0         & 0         & 0\tabularnewline
\hline

\end{tabular}
}
\end{adjustbox}
\vspace{-2ex}
\caption{Delay of writing packets to the VPN tunnel under four different writing schemes.}
\vspace{-1ex}
\label{tab:writeoverhead}
\end{table}

According to Table~\ref{tab:writeoverhead}, the queueWrite scheme
performs much better than the directWrite scheme.  Among a total of
1,244 samples in the directWrite testing, we encounter 42 large
writing overheads (i.e., those larger than 1ms).  The corresponding
result for the queueWrite testing is only 14 out of 2,161 samples.  In
particular, there are five extremely large overheads (i.e., those
larger than 5ms) in the directWrite samples, two of which are even
over 20ms.
While there are still 14 overheads of 1$\sim$5ms for queueWrite, they do not affect the performance of \texttt{MainWorker},
because they are performed by the dedicated \texttt{TunWriter} thread.

Although the queueWrite scheme significantly reduces the writing
overhead, it introduces the overhead of packet enqueuing.  We find
that a traditional enqueuing scheme, denoted by oldPut, has
large overheads.  Among the 810 oldPut samples in
Table~\ref{tab:writeoverhead}, 47 have an overhead larger than 1ms.
Our testing shows that most of the overheads between 1$\sim$5ms are
due to the queue's \texttt{wait-notify} delay. When there are no
packets in the queue, \texttt{TunWriter} goes to sleep by calling
\texttt{queue.wait()} and is woken up by \texttt{queue.notify()}.  We
design a new enqueuing algorithm, denoted by newPut, to
mitigate such delays.  The basic idea is to let \texttt{TunWriter}
perform more rounds of queue checking before going to \texttt{wait()}.
Specifically, we design a sleep counter to systemize this process:
\begin{compactitem}
  \item The counter is initialized to 0 and is reset to 0 each time
    being woken up from \texttt{wait()}.
  \item When there are no packets in the queue, the counter increments
    for every round of checking and decrements (e.g., dividing by 2)
    whenever detecting a nonempty queue.
  \item \texttt{TunWriter} sleeps only when the counter reaches a
    threshold.
\end{compactitem}

The newPut column in Table~\ref{tab:writeoverhead} shows the
effectiveness of our algorithm.  Out of the 5,321 samples, only four
contain 1$\sim$5ms overheads.  Compared with the oldPut scheme, the
percentage of large overheads drops from 5.69\% to only 0.075\%.  It
is worth noting that the remaining two large overheads of 5$\sim$10ms
are likely due to thread competition.  We also observe that such
competition effect is significantly reduced, because the enqueuing
operation (at the microsecond level) is much faster than tunnel
writing (at the 0.1ms level).

\vspace{-2.5ex}
\subsubsection{Dispatching of Socket Packets}
\label{sec:socketwrite}
\vspace{-1.3ex}
When \name relays packets to the external connection, a delay overhead
which could be up to several milliseconds comes from the
\texttt{VpnService.protect(socket)} method \cite{VPNprotect}. Before establishing socket connections with remote
app servers, \name must call the \texttt{protect(socket)} method to
ensure that the socket packets will be sent directly to the underlying
network. Without this method, the socket packets will be directed back
to the VPN tunnel, thus creating a data loop.

Our solution is to replace the socket-wide \texttt{protect()} API with
the application-wide \texttt{addDisallowedApplication()} API.  By
adding \name into the list of VPN-disallowed applications, we do not
need to invoke \texttt{protect(socket)} for each socket client.
Moreover, since \name just needs to call
\texttt{addDisallowedApplication(mopeye)} once, the call is best
invoked during the initialization of \name to avoid impact on
\texttt{MainWorker}.  The limitation of this solution is that
\texttt{addDisallowedApplication()} is newly introduced in Android
5.0.  For older versions, \name still has to call
\texttt{protect(socket)}. Our mitigation method is to put
\texttt{protect(socket)} in each \texttt{socket-connect} thread. In
this way, only the performance of the SYN packet will be affected but
not the subsequent data.
Furthermore, this issue will be of less importance as more devices are upgraded to Android 5.0+, currently with over 60\% of devices~\cite{dashboards}.

\section{Evaluation}
\label{sec:evaluate}
\vspace{-2ex}


In this section, we present two sets of evaluation results.
The first is on the measurement accuracy and overhead of \name, and the second is a set of crowdsourcing measurement results from 2,351 active users over nine months.

\vspace{-2ex}
\subsection{Measurement Accuracy and Overhead}
\label{sec:validate}
\vspace{-1ex}

\subsubsection{Measurement Accuracy}
\label{sec:accuracy}

The first evaluation we perform is on the accuracy of RTT measurement
of \name.  In addition to the standalone measurement, we also compare
\name with MobiPerf~v3.4.0 (the latest version at the time of our
evaluation), which makes active network measurements using the state-of-the-art Mobilyzer
library~\cite{Mobilyzer15}. For a fair comparison, we use MobiPerf's
HTTP ping measurement~\cite{wchli15} because, like \name, it also uses
SYN-ACK for the RTT measurement. For each destination, we use its raw
IP address instead of the domain name so that MobiPerf's accuracy
will not be interfered by DNS queries. Moreover, each result is
presented by the mean of ten independent runs (MobiPerf does not
provide detailed results of each run). We also run \texttt{tcpdump} to
provide the reference measurement results.

\begin{table}[h!]
\vspace{-1ex}
\begin{adjustbox}{center}
\scalebox{
0.78
}{
\begin{threeparttable}
\begin{tabular}{ |c | c | c | c | c | c | c | c | c |}

\hline
\multirow{3}{*}{Destinations} & \multicolumn{3}{c|}{\name (mean, in ms)} & \multicolumn{3}{c|}{MobiPerf (mean, in ms)}  \tabularnewline
\cline{2-7}
& \texttt{tcp}   & Mop    & \multirow{2}{*}{$\delta$} &  \texttt{tcp}    & Mobi    &  \multirow{2}{*}{$\delta$}  \tabularnewline
& \texttt{dump}   & Eye*  &   & \texttt{dump}    & Perf    &  \tabularnewline
\hline
\hline

\multirow{2}{*}{Google} & 4.26 &  4    & \textbf{0}      & 4.29  & 16.4  & \textbf{12.11}   \tabularnewline
\multirow{2}{*}{(\texttt{216.58.221.132})} & 4.47 &  5.5  & \textbf{1.03} & 4.35  & 18.5  & \textbf{14.15} \tabularnewline
                      & 5.32 &  5    & \textbf{0}    &  4.85 & 18    & \textbf{13.15}   \tabularnewline
\hline

\multirow{2}{*}{Facebook}& 36.55 & 37   & \textbf{0.45}    & 36.39 &  59.5  & \textbf{23.11}  \tabularnewline
\multirow{2}{*}{(\texttt{31.13.79.251})}  & 36.55 & 37   & \textbf{0.45}  & 36.72 & 55.2  & \textbf{18.48} \tabularnewline
                      & 38.54 & 38.5   & \textbf{0}   & 46.10 & 63.2  & \textbf{17.10} \tabularnewline
\hline

\multirow{2}{*}{Dropbox}& 284.85 & 284.5   & \textbf{0}     & 361.76 &  409.7  & \textbf{47.94}  \tabularnewline
\multirow{2}{*}{(\texttt{108.160.166.126})} & 390.94 & 391  & \textbf{0.06}    & 388.94 & 411.5  & \textbf{22.56} \tabularnewline
                      & 513.78 & 513.5   & \textbf{0}   & 395.87 & 475.2  & \textbf{79.33} \tabularnewline
\hline

\end{tabular}
\begin{tablenotes}
\item [*] \small We round \name's $\mu$s-level results to $ms$-level, e.g., 4.135ms to 4ms.
\end{tablenotes}
\end{threeparttable}
}
\end{adjustbox}
\vspace{-3ex}
\caption{\small Measurement accuracy of \name and MobiPerf.}
\vspace{-2ex}
\label{tab:VSmobiperf}
\end{table}

Table~\ref{tab:VSmobiperf} presents three sets of results for Google,
Facebook, and Dropbox, which experience RTTs on different scales.
The differences between the RTT measurement of MopEye/MobiPerf and that of
\texttt{tcpdump} are denoted by $\delta$. The results clearly show
that \name has a much better accuracy than MobiPerf---\name's
measurement deviates from that of \texttt{tcpdump} by at most 1ms,
whereas MobiPerf's deviations range from 12ms to 79ms.
By assessing MobiPerf's code\footnote{\url{http://tinyurl.com/PingTask}, where HTTP ping starts from the line 438.}, we identify three factors responsible for \name's higher accuracy, including using the low-level socket call and the nanosecond-level timestamp method, and most importantly, putting the timing function just before and after the socket call.
We refer interested readers to our previous poster version~\cite{MopEyePoster15} for more details.


\subsubsection{Measurement Overhead}
\label{sec:latency}
\vspace{-1ex}
To measure the overhead introduced by \name, we first measure the
additional delay introduced to the connection establishment and data transmission in
other apps when \name is running.  For the connection time, we
implement a simple tool that invokes \texttt{connect()} to measure the
time taken with and without \name.  For data packets, we use the
popular Ookla Speedtest app~\cite{speedtest_android} to measure the
latency.  Both experiments are repeatedly
executed on a Nexus~4 running Android~5.0.  With a 95\% confidence
interval, the mean delay overhead of a round of SYN and SYN/ACK packets is
3.26$\sim$4.27ms and that of data packets is 1.22$\sim$2.18ms.
Considering that the median of all 714,675 LTE RTTs in our dataset is 76ms, the delay overhead is acceptable.

Another important metric is the download and upload throughput
overhead.  We compare \name with Haystack~\cite{Haystack15} v1.0.0.8
(the latest version at the time of our evaluation),
which uses the \texttt{VpnService} API to detect privacy leaks in app traffic.
For a fair comparison, we do not enable Haystack's TLS traffic analysis for all
its experiments.  We use the Ookla Speedtest app as the reference tool
to measure the throughput with and without \name/Haystack.  All three
experiments are repeatedly conducted in a dedicated WiFi network
which provides very strong signal strength and stable throughput at
around 25Mbps for both download and upload links.

\begin{table}[t!]
\vspace{-3ex}
\begin{adjustbox}{center}
\scalebox{
0.83
}{
\begin{tabular}{ |c | c || c | c || c | c| }

\hline
Throughput  & Baseline  & MopEye  & $\Delta$  & Haystack  & $\Delta$ \tabularnewline
\hline
\hline

Download    & 24.47    & 24.01    & \textbf{0.46}    & 20.19    & \textbf{4.28} \tabularnewline
\hline
Upload      & 25.97    & 25.08    & \textbf{0.89}    & 6.79     & \textbf{19.18} \tabularnewline
\hline

\end{tabular}
}
\end{adjustbox}
\vspace{-3ex}
\caption{\small The download and upload throughput overhead of \name and Haystack.}
\vspace{-3ex}
\label{tab:throughput}
\end{table}

Table \ref{tab:throughput} presents the throughput results with $\Delta$ denoting the
difference from our baseline using Speedtest.
The results clearly show that \name achieves a much better throughput
performance than Haystack. \name's throughput deviates from the
baseline by less than 1Mbps, whereas that for Haystack ranges from 4Mbps
(for the download link) to 19Mbps (for the upload link).  In
particular, we find that Haystack's throughput degrades significantly
(e.g., 11.63Mbps for the download and 3.74Mbps for the upload) if we
do not restart it for the next run.  Therefore, in order to obtain the
Haystack results in Table~\ref{tab:throughput}, we reset
Haystack's VPN interface before each test.  We attribute our superior results to the
major challenges addressed in \mysec\ref{sec:implement}.

%
%
%
%
%

\vspace{-2ex}
\subsubsection{Resource Consumption Overhead}
\label{sec:resource}
\vspace{-1ex}

We now summarize the resource consumption overhead of \name and Haystack with a Nexus~6 playing a high-definition YouTube video for around one hour.
According to Table \ref{tab:resourceoverhead}, \name's resource consumption overhead is lower than that of Haystack in terms of CPU, battery, and memory.
In particular, the CPU overhead with Haystack is over 9\%, mainly because Haystack has to keep executing the VPN \texttt{read()} regardless there are app packets to be relayed or not.
Moreover, we argue that the 1\% battery overhead of \name is not contributed only by \name, because, with \name enabled, YouTube is no longer considered using the network interface by the system battery benchmark.

\begin{table}[t!]
\vspace{-3ex}
\begin{adjustbox}{center}
\scalebox{
0.9
}{
\begin{tabular}{ |c | c | c | c| }

\hline
\multirow{2}{*}{Scenario}   & \multirow{2}{*}{Resource} & \multicolumn{2}{c|}{Overhead} \tabularnewline
\cline{3-4}
                            &                       & \name     & Haystack \tabularnewline
\hline
\hline


Playing a 58-minute         & CPU       & 2.74\%        & 9.56\% \tabularnewline
\cline{2-4}
high-definition (1080p)     & Battery   & 1\%           & 2\% \tabularnewline
\cline{2-4}
YouTube video               & Memory    & 12MB          & 148MB \tabularnewline
\hline

\end{tabular}
}
\end{adjustbox}
\vspace{-3ex}
\caption{\small The resource overhead of \name and Haystack.}
\vspace{-3ex}
\label{tab:resourceoverhead}
\end{table}

\vspace{-2.5ex}
\subsection{Crowdsourcing Measurement Results}
\label{sec:result}
\vspace{-1ex}

Our \name deployment on Google Play has attracted 4,014 user installs from 126 countries since May 2016.
In this section, we first describe the dataset used in this paper and then present our measurement analysis to underline the value of \name's opportunistic per-app measurement.

\vspace{-2.5ex}
\subsubsection{Dataset Statistics}
\label{sec:dataset}
\vspace{-1ex}

By deploying \name for over ten months, to the best of our knowledge, we have collected the \textit{first} large-scale per-app measurement dataset.
Our analysis in this paper is based on the \name data received between its launch on 25 May 2016 and 3 January 2017.
Our dataset covers a wide spectrum of devices, countries, and apps, and includes over 5 million RTT measurements.

\begin{figure}[t!]
\vspace{-1ex}
\begin{adjustbox}{center}
  \subfigure[By user.] {
    \label{fig:PerUserNumBar}
    \includegraphics[width=0.27\textwidth]{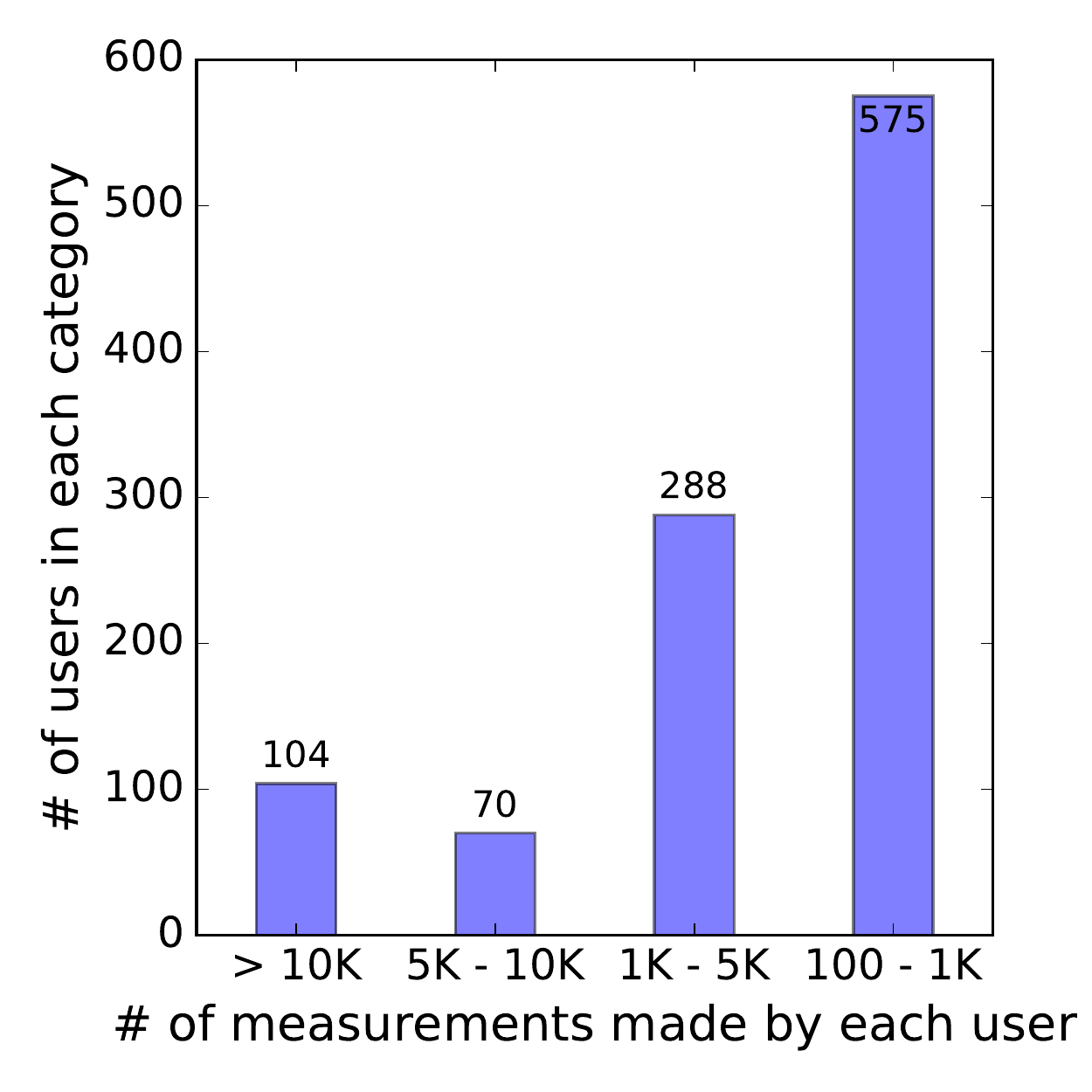}
  }
  \hspace{-2.5ex}
  \subfigure[By app.] {
	\label{fig:PerAppNumBar}
    \includegraphics[width=0.27\textwidth]{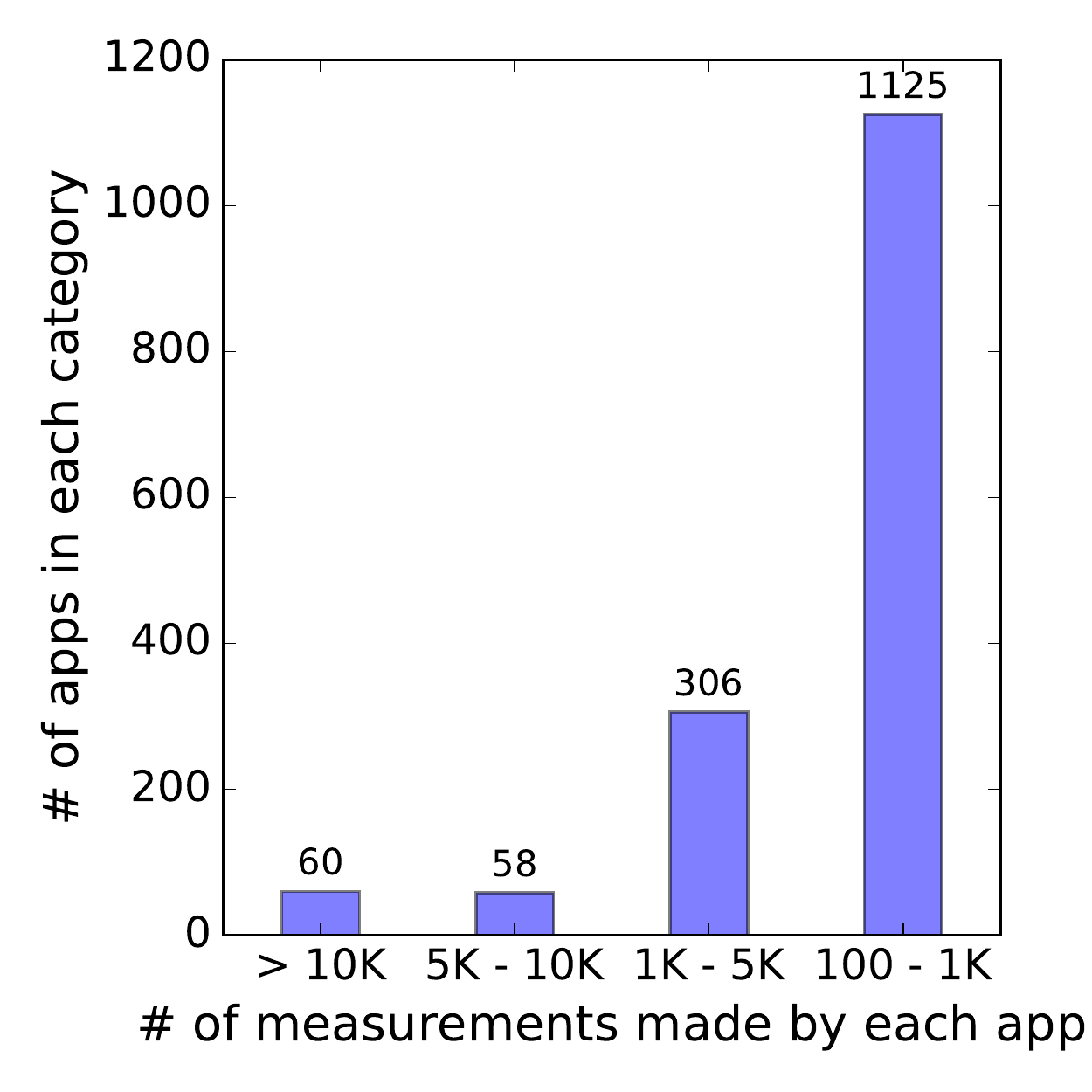}
  }
\end{adjustbox}
\vspace{-4ex}
\caption{\small Number of measurements performed by each user/app that contribute to at least 100 measurements.}
\vspace{-3.5ex}
\label{fig:}
\end{figure}

\begin{compactitem}
%
%
\item [\bf{User/Device coverage:}]
The dataset includes a total of 2,351 devices that performed at least one measurement.
\myfig~\ref{fig:PerUserNumBar} shows the number of measurements performed by 1,037 devices each of which conducted at least 100 measurements.
Although most of them are in the range of 100--1K, 462 of them (45\%) contribute from 1K to more than 10K measurements each. This shows a significant number of consistently active users.
Moreover, these user devices cover 922 different phone models, manufactueres of which include Samsung, HTC, LG, Motorola, Huawei, XiaoMi, and others. This evidences that \name can support a wide range of Android phones in the market.

\item [\bf{Country distribution:}]
Users in our dataset come from 114 countries worldwide.
\myfig~\ref{fig:UserCountry} shows the distribution of the top 20 user countries, including the United States (790 users), United Kingdom (116 users), India (70 users), and Italy (68 users).
Moreover, \myfig~\ref{fig:location} plots 6,987 geographical locations where the \name measurements were conducted. The figure visually shows that our dataset covers a large populated area, notably the North America, Europe, India, coastal regions of South America, Southeast Asia, and the Pacific Rim.


\item [\bf{Applications measured:}]
This dataset includes measurement on 6,266 apps.
\myfig~\ref{fig:PerAppNumBar} shows the distribution of the number of RTT measurements performed by each app that contributes at least 100 measurements, with a total of such 1,549 apps. Similar to \myfig~\ref{fig:PerUserNumBar}, most of them contribute 100--1K measurements, and 424 of them have between 1K and more than 10K measurements. The most popular (in terms of the number of times being measured) apps include social networking apps such as Facebook, Instagram, and WeChat, and system built-in apps such as YouTube and Google Play.

\begin{figure}[t!]
\vspace{-2ex}
\begin{adjustbox}{center}
\includegraphics[width=0.35\textwidth]{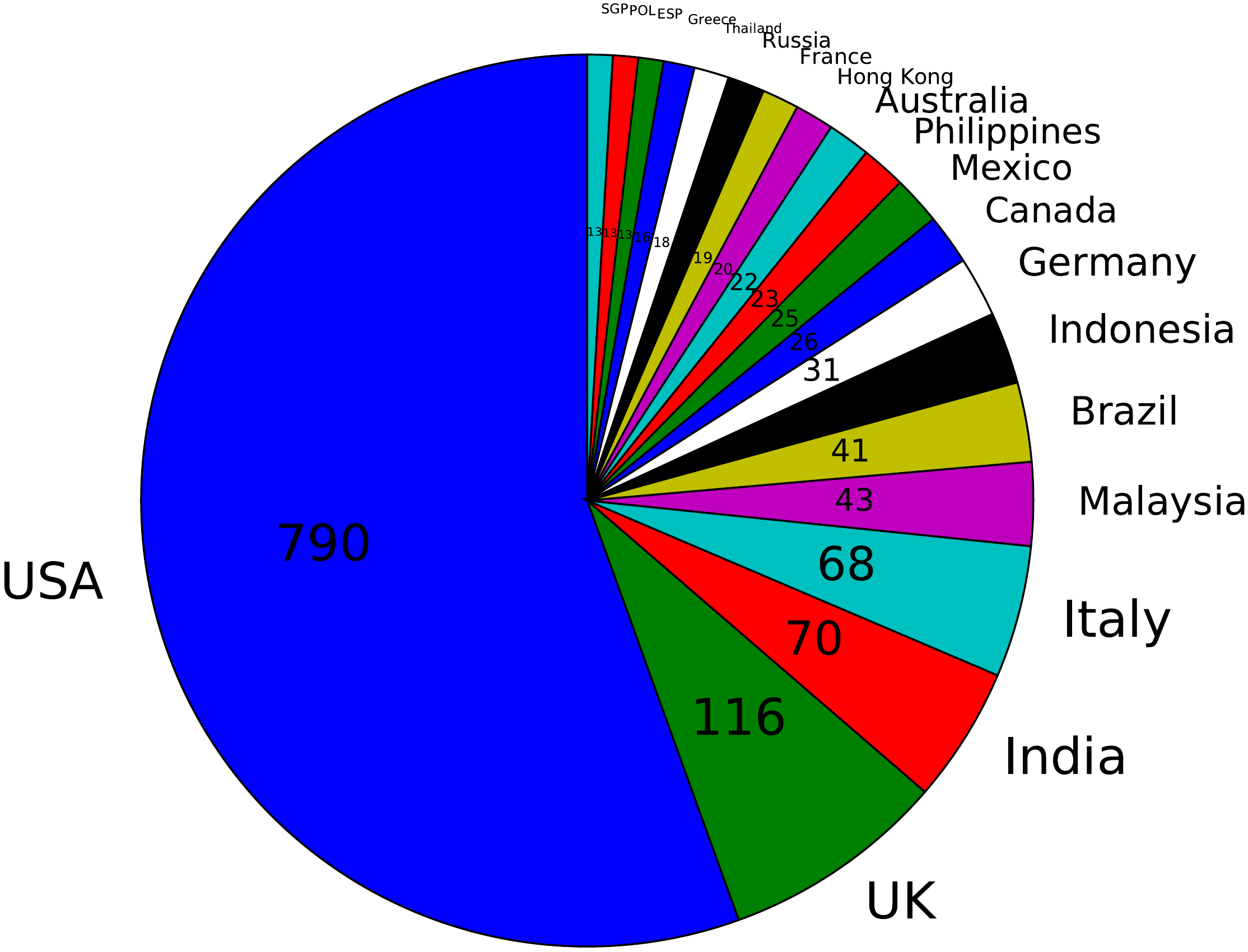}
\end{adjustbox}
\vspace{-4ex}
\caption{\small Distribution of the top 20 \name user countries.}
\vspace{-2ex}
\label{fig:UserCountry}
\end{figure}

\begin{figure}[t!]
\begin{adjustbox}{center}
\includegraphics[width=0.42\textwidth]{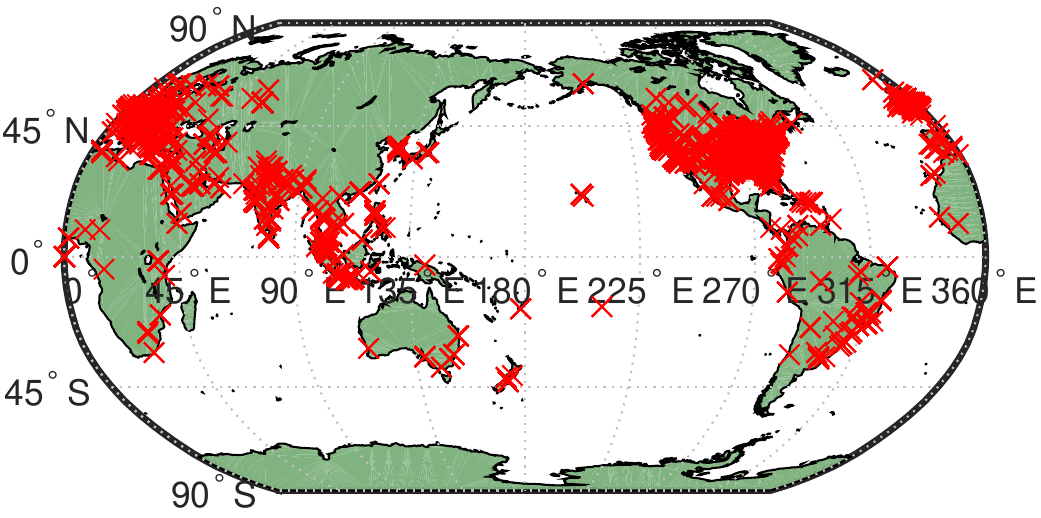}
\end{adjustbox}
\vspace{-4ex}
\caption{\small Locations of conducting the \name measurement.}
\vspace{-4ex}
\label{fig:location}
\end{figure}

%
%
\item [\bf{Measurements collected:}]
The dataset contains a total number of 5,252,758 RTT measurements.
Among them, 3,576,931 are measurements for TCP connections used by the apps, and the remaining 1,675,827 are for DNS measurements.
Altogether they cover 106,182 destination IP addresses, 35,351 destination server domains, 2,427 destination server ports, and 943+ DNS servers.
The most accessed domain is \texttt{graph.facebook.com} with 142,873 connections.
\end{compactitem}

\vspace{-2ex}
\subsubsection{Per-app Measurement Analysis}
\label{sec:tcpresult}
\vspace{-1ex}

We now present the 3,576,931 per-app measurement results, which characterize the network performance experienced by different apps under different network types and ISPs in the wild.
We envision ways of using the analysis results to improve the mobile network performance. For example, we reported our measurement results of WeChat to help Tencent (developer of the WeChat app) solve a misconfiguration problem~\cite{MopEyePoster15}.

%
%

\begin{figure}[t!]
\vspace{-4ex}
  \begin{adjustbox}{center}
  \subfigure[All apps' raw RTTs.] {
    \label{fig:rawAppRTT}
    \includegraphics[width=0.26\textwidth]{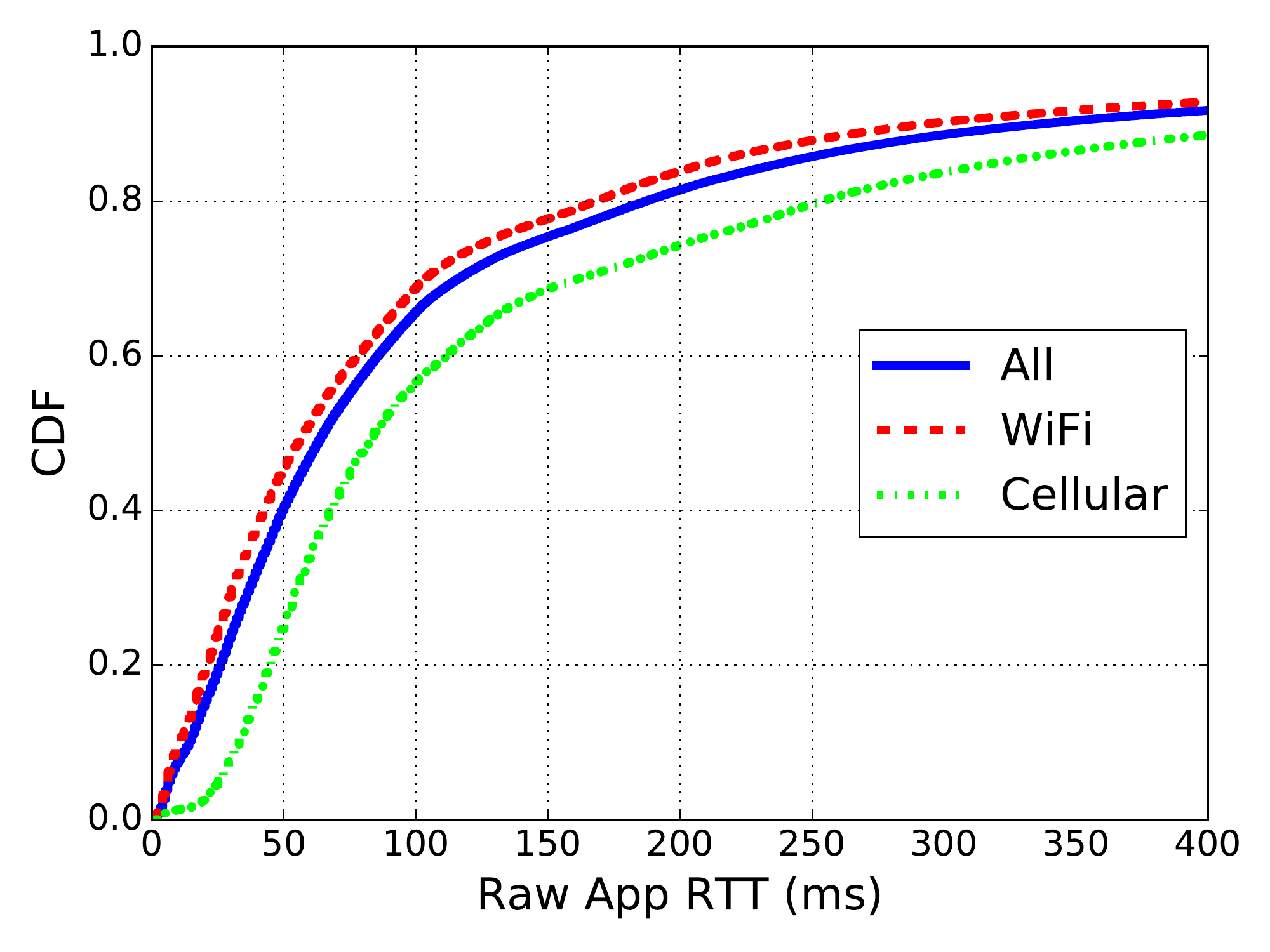}
  }
  \hspace{-2.3ex}
  \subfigure[Top 424 apps' median RTTs.] {
	\label{fig:medAppRTT}
    \includegraphics[width=0.26\textwidth]{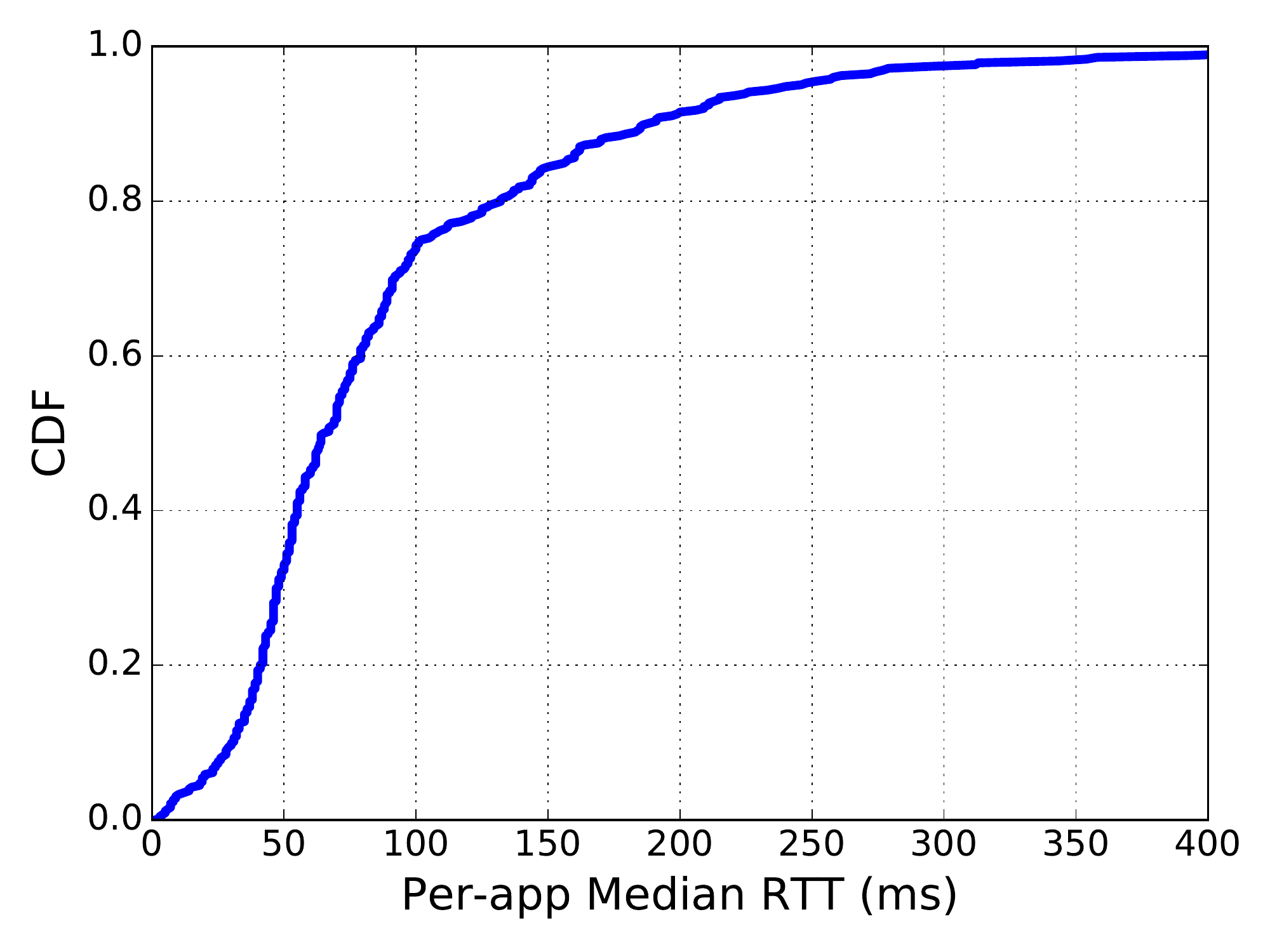}
  }
  \end{adjustbox}
\vspace{-4ex}
\caption{\small CDF plots of apps' raw RTTs and median RTTs.}
\vspace{-3ex}
\label{fig:overallAppRTT}
\end{figure}

\textbf{Overall results.}
We first present the overall app performance in our dataset by plotting the distribution of apps' raw and median RTTs in \myfig~\ref{fig:overallAppRTT}.
\myfig~\ref{fig:rawAppRTT} shows the CDF plot of all 6,266 apps' raw RTTs, in which we further distinguish between WiFi and cellular access.
Overall, the performance experienced by mobile users is good with a median RTT of 65ms (i.e., the value at the 0.5 line in \myfig~\ref{fig:rawAppRTT}).
Moreover, $\sim$40\% of the RTTs are below 50ms and $\sim$60\% of the RTTs are below 100ms.
However, we can still observe $\sim$20\% of them suffering from relatively long RTTs ($>$200ms), and $\sim$10\% at exceedingly long RTT ($>$400ms). In this dataset, WiFi shows superior performance than that on cellular networks. The median RTTs for WiFi, cellular networks (including 2G, 3G, and LTE), and LTE alone are 58ms, 84ms, and 76ms, respectively.

\myfig~\ref{fig:medAppRTT} plots the median RTT distribution of 424 apps that have more than 1K measurements each (see \myfig~\ref{fig:PerAppNumBar}). We choose the median over the mean value because the median is less affected by RTT outliers.
The dataset also has enough measurements for each app, making the median a reliable measure.
The figure shows that more than 70\% of the apps experience less than 100ms in their RTTs.
However, there are $\sim$10\% of the apps suffering from more than 200ms of RTT.

\textbf{Representative apps' performance.}
We next study the network performance of representative apps that are frequently used in our daily life.
Table~\ref{tab:topapp} lists 16 such apps in five categories.
For each app, we present its total number of RTT measurements and the median RTT.
Most of these apps exhibit very good network performance.
For example, Instagram, WeChat, Google Play Store, YouTube, and Amazon have a median RTT below 60ms.
We also notice that the median RTT of Whatsapp is larger than 100ms. Next we present two case studies in more depth.

\begin{table}[t!]
\vspace{-2.8ex}
\begin{adjustbox}{center}
\scalebox{
0.76
}{
\begin{tabular}{ |c | c | c | c | }

\hline
Category    & Apps          & \# RTT    & Median RTT\tabularnewline 
\hline
\hline

\multirow{4}{*}{Social}
            & Facebook      & 215,769   & 61ms      \tabularnewline 
            & Instagram     & 38,640    & 50.5ms    \tabularnewline 
            & Weibo         & 28,905    & 43ms      \tabularnewline 
            & Twitter       & 11,407    & 56ms      \tabularnewline 
\hline

            & WeChat        & 61,804    & 36ms      \tabularnewline 
Commu-      & Facebook Messenger  & 42,408    & 42ms      \tabularnewline 
nication    & Whatsapp      & 32,372    &\red{133ms}\tabularnewline 
            & Skype         & 16,264    & 76ms      \tabularnewline 
\hline

\multirow{4}{*}{Google}
            & Google Play Store    & 100,115   & 48ms      \tabularnewline 
            & Google Play services & 60,805    & 37ms      \tabularnewline 
            & Google Search & 35,858    & 45ms      \tabularnewline 
            & Google Map    & 19,996    & 38ms      \tabularnewline 
\hline

\multirow{2}{*}{Video}
            & YouTube       & 99,895    & 32ms      \tabularnewline 
            & Netflix       & 28,302    & 33ms      \tabularnewline 
\hline


\multirow{2}{*}{Shopping}
            & Amazon        & 18,313    & 59ms      \tabularnewline 
            & Ebay          & 16,114    & 70ms      \tabularnewline 
\hline


\end{tabular}
}
\end{adjustbox}
\vspace{-3ex}
\caption{\small Network performance of 16 representative apps.}
\vspace{-4ex}
\label{tab:topapp}
\end{table}

%
%
\textbf{Case 1: The vast majority of *.whatsapp.net domains do not perform well in many networks.}
Whatsapp employs a total of 334 \url{whatsapp.net} domains as its server domains, but the median RTT of all these domain traffic is as high as 261ms.
Specifically, the median RTTs for all, except three, are larger than 200ms.
The median RTTs for those three domains (starting with \texttt{mme}, \texttt{mmg}, or \texttt{pps}) are less than 100ms. According to our analysis, the three domains are deployed in the Facebook CDN, whereas the other 331 domains are with SoftLayer Technologies, a server hosting provider.
Furthermore, we analyze the median RTTs on these 331 \url{whatsapp.net} domains in 20 most accessed networks (11 WiFi and 9 LTE networks) that have at least 100 measurements each.
The results show that only two networks can achieve less than 100ms of RTT (77.5ms for a WiFi network and 56ms for the Verizon 4G network), six networks in the 100--200ms interval, eight networks in between 200ms and 300ms, and four networks with RTTs over 300ms.
Moreover, our manual Ping tests from Singapore and Hong Kong to those domains report a latency of $\sim$250ms.
All of the above show that there is much room for Whatsapp to improve their \url{whatsapp.net} network performance.

%
%
\textbf{Case 2: Jio, India's largest 4G ISP, fails to provide acceptable performance to many app domains.}
In the course of analyzing the Whatsapp case, we find that Jio provides poor performance to many app server domains.
Among all the ten 4G ISPs with more than 10K measurements, Jio is the only one that has a median RTT larger than 100ms.
The median RTT of its 76,717 RTT measurements is as high as 281ms.
Considering that the median RTT of its DNS measurements is only 59ms, the root cause lies very likely in its LTE core network.
Moreover, our analysis of 115 domains (that have 100+ measurements each) in Jio finds that only 19 domains' median RTTs are less than 100ms, whereas the median RTTs of 67 domains are over 200ms, 57 domains over 300ms, and 24 domains even over 400ms.
We further confirm that Jio's poor performance is not due to the performance of the app servers.
It is because out of the 71 domains that have 100+ measurements each in both Jio and non-Jio LTE networks, 
63 of them have much better latency (138ms less than Jio on average) with non-Jio LTE networks.


%
%


\vspace{-2ex}
\subsubsection{DNS Measurement Analysis}
\label{sec:dnsresult}
\vspace{-1ex}

Next we analyze the 1,675,827 DNS measurements received from 943+ WiFi and cellular DNS servers.

\begin{figure}[t!]
\vspace{-3.5ex}
  \begin{adjustbox}{center}
  \subfigure[All results.] {
    \label{fig:AllDNS}
    \includegraphics[width=0.26\textwidth]{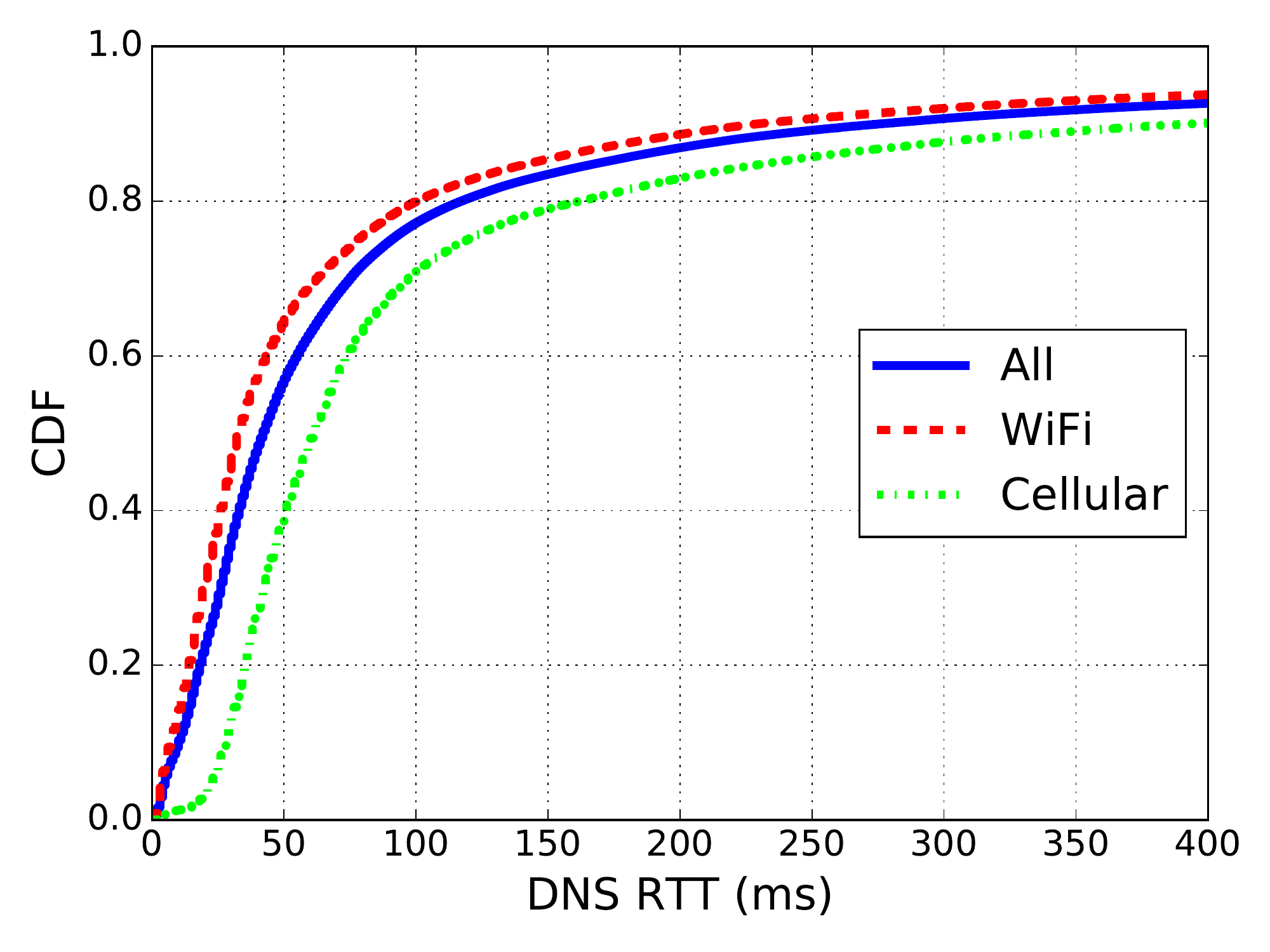}
  }
  \hspace{-2.3ex}
  \subfigure[Cellular results.] {
	\label{fig:CellDNS}
    \includegraphics[width=0.26\textwidth]{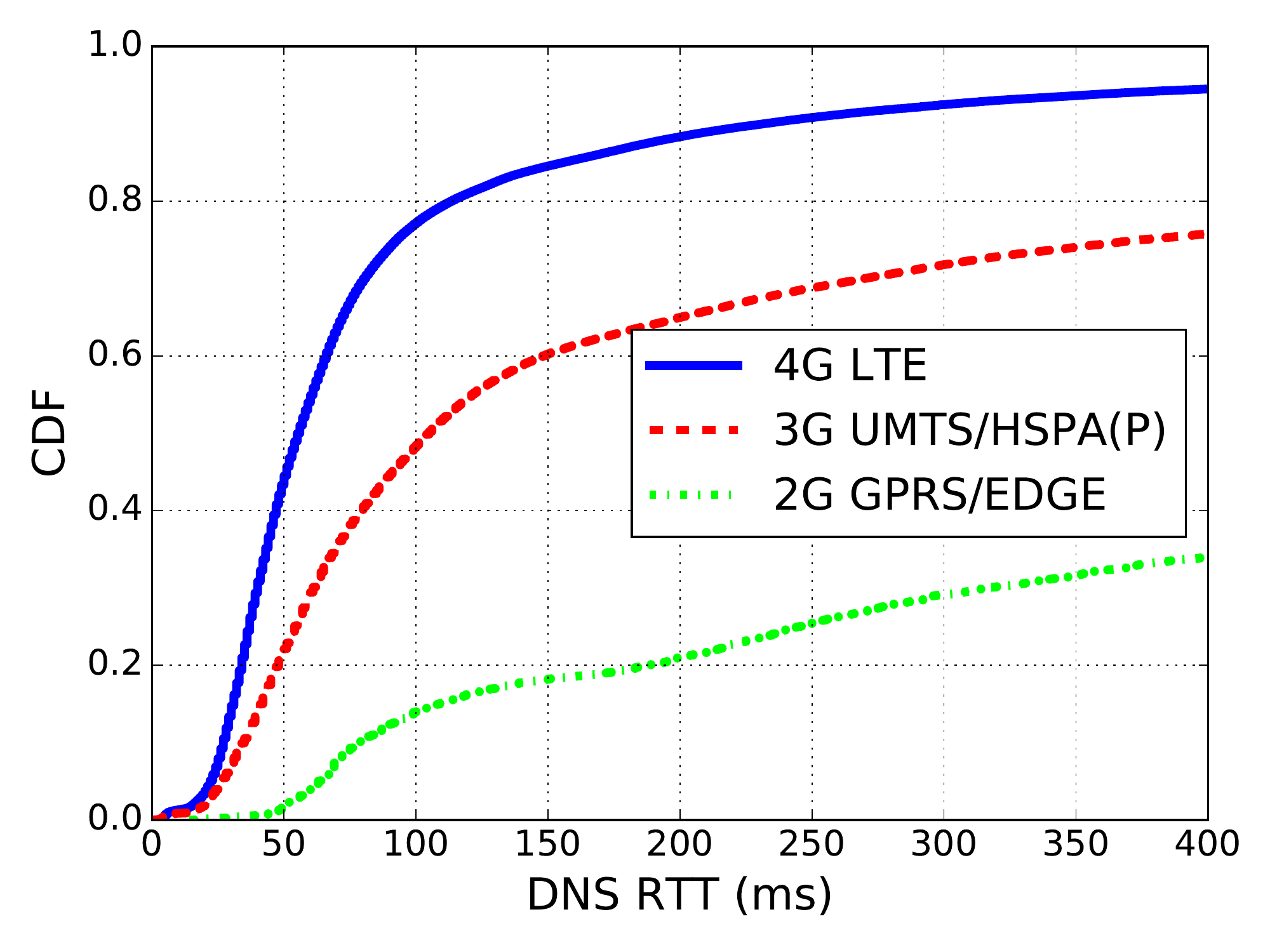}
  }
  \end{adjustbox}
\vspace{-4ex}
\caption{\small CDF plots of DNS measurement results.}
\vspace{-4ex}
\label{fig:}
\end{figure}

\textbf{Overall results.}
\myfig~\ref{fig:AllDNS} shows the CDF plot of all measured DNS RTTs.
According to the overall distribution, the DNS performance for mobile networks in the wild is good with a median of 42ms, and around 80\% of DNS RTTs are less than 100ms.
The DNS RTTs are in fact much better than the per-app performance by comparing \myfig~\ref{fig:AllDNS} with \myfig~\ref{fig:rawAppRTT}. For example, 80\% of per-app RTTs are less than 200ms, two times higher than DNS.
This is probably because ISPs usually deploy local DNS servers.
Additionally, we notice that WiFi's DNS RTTs are consistently lower than the overall results with a median of only 33ms; whereas that of cellular networks is 61ms.
This indicates that the first-hop performance of WiFi is generally better than cellular networks.

We plot the detailed results for 2G, 3G, and 4G cellular networks in \myfig~\ref{fig:CellDNS}.
The CDF plots show clearly the performance difference among the three.
More specifically, the median DNS RTT of 4G is 56ms; whereas that of 3G and 2G are as high as 105ms and 755ms, respectively.
Most of the devices in our measurement use 4G---around 80\% of DNS RTTs come from 4G.  
This also explains why the CDF plot for 4G DNS RTTs is close to that of all cellular RTTs.

\textbf{Major 4G ISPs' DNS performance.}
We now take a closer look at the DNS performance of major 4G ISPs.
Table~\ref{tab:dnsISP} lists the performance of 15 LTE operators that have most DNS RTTs in our dataset.
First, we notice that there is no clear correlation between the country and DNS performance.
For example, the performance of most American ISPs, three Hong Kong ISPs, and two Malaysia ISPs are similar.
Second, the majority of 4G ISPs provide good DNS performance with the median RTTs in 40-60ms.
The only three outliers are the good-performer Singtel, and the poor-performers Cricket and U.S. Cellular.
To gain a better understanding, we further study these three ISPs along with Verizon, a representative of other ISPs.

\begin{table}[t!]
\vspace{-1ex}
\begin{adjustbox}{center}
\scalebox{
0.74
}{
\begin{tabular}{ |c | c | c | c | }

\hline
ISP Name    & Country   & \# RTT    & Median RTT\tabularnewline 
\hline
\hline

Verizon     & America   & 80,227    & 46ms      \tabularnewline 
\hline
Jio 4G      & India     & 52,397    & 59ms      \tabularnewline 
\hline
AT\&T       & America   & 51,421    & 53ms      \tabularnewline 
\hline
Singtel     & Singapore & 34,609    & \blue{27ms}      \tabularnewline 
\hline
Boost Mobile& America   & 21,854    & 50ms      \tabularnewline 
\hline
Sprint      & America   & 20,878    & 51ms      \tabularnewline 
\hline
3           & HK (China)& 14,354    & 53ms      \tabularnewline 
\hline
MetroPCS    & America   & 13,282    & 60ms      \tabularnewline 
\hline
T-Mobile    & America   & 9,084     & 45ms      \tabularnewline 
\hline
CMHK        & HK (China)& 5,820     & 50ms      \tabularnewline 
\hline
Celcom      & Malaysia  & 4,120     & 56ms      \tabularnewline 
\hline
CSL         & HK (China)& 3,099     & 61ms      \tabularnewline 
\hline
Cricket     & America   & 2,822     & \red{93ms}\tabularnewline 
\hline
Maxis       & Malaysia  & 2,419     & 40ms      \tabularnewline 
\hline
U.S. Cellular& America  & 1,988     & \red{76ms}\tabularnewline 
\hline

\end{tabular}
}
\end{adjustbox}
\vspace{-3ex}
\caption{\small DNS performance of 15 LTE 4G operators.}
\vspace{-2ex}
\label{tab:dnsISP}
\end{table}

\begin{figure}[t!]
\begin{adjustbox}{center}
\includegraphics[width=0.26\textwidth]{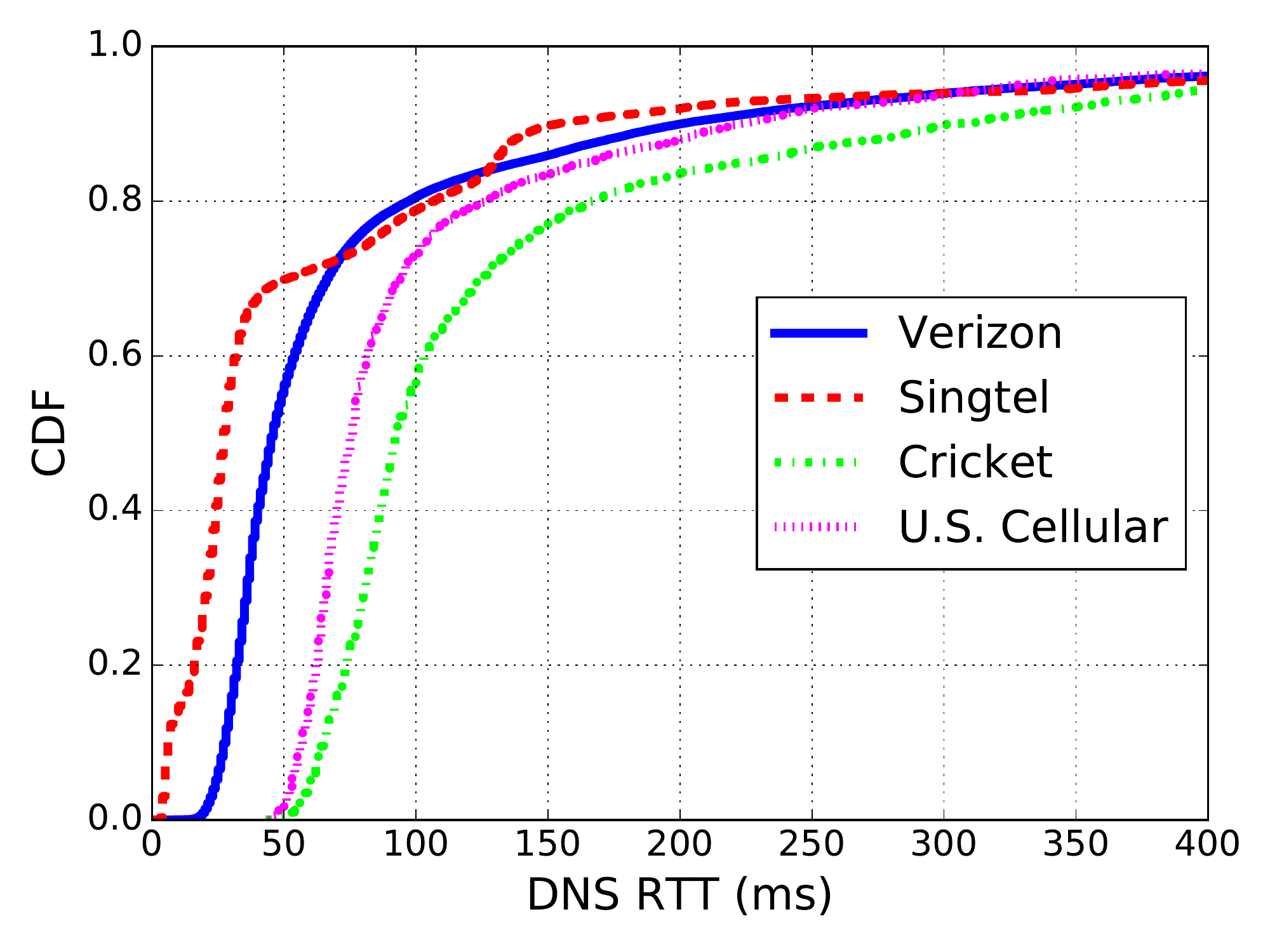}
\end{adjustbox}
\vspace{-5ex}
\caption{\small CDF plots for DNS performance of four LTE ISPs.}
\vspace{-3ex}
\label{fig:DNScaseCDF}
\end{figure}

\myfig~\ref{fig:DNScaseCDF} presents the DNS RTT distribution of the four selected ISPs with the Verizon plot as the baseline.
The plots show that Singtel has an outstanding first-hop performance with 5,084 DNS RTTs less than 10ms (14.7\% of its total RTTs), whereas Verizon has less than 1\% of its DNS RTTs below 10ms. 
This is mainly because Singtel has deployed the latest upgrade of LTE, Tri-band 4G+~\cite{Latest4G}, countrywide~\cite{Singtel4GHistory}.
On the other hand, the DNS performance of Cricket and U.S. Cellular clearly is worse than the baseline.
In particular, the minimum RTTs of Cricket and U.S. Cellular are around 43ms, much higher than the best performance of Singtel and Verizon.
They are probably using the pre-4G or near-4G implementations, because we find that around half of their DNS RTTs (64\% of Cricket and 45\% of U.S. Cellular) are still from non-LTE networks.

%


\vspace{-2ex}
\begin{framed}
\vspace{-1.5ex}
\small
\noindent
\textbf{Key Takeaway:} 
\name enables a large-scale deployment of per-app measurements in the wild, which help understand and diagnose the network quality of app providers and mobile networks at different granularity.
\vspace{-1.5ex}
\end{framed}
\vspace{-4ex}

\vspace{-1ex}
\section{Related Work}
\label{sec:related}
\vspace{-3ex}

Many measurement tools have been proposed to understand mobile network performance. They could be classified into crowdsourcing measurement apps (e.g., \cite{3GTest10, 4GTest12, WiFiLTE14, Mobilyzer15}) and controlled testbeds (e.g., \cite{ProfileDroid12, NornetEdge14, QoEDoctor14}).
They study 3G/4G networks' RRC (Radio Resource Control) state dynamics~\cite{RRC3G10, 4GTest12, RRCState14}, analyze the behaviors of cellular networks~\cite{First10, IdentifyBehavior11, Bufferbloat12, LTE13}, measure mobile network performance and reliability~\cite{Mobilyzer15, NornetEdge14, IMC14, PktLoss15, CellDNS14}, and perform other measurements~\cite{wchli15,CoNEXT16}.
\name belongs to the domain of crowdsourcing measurements.
Using the \texttt{VpnService} API to perform passive network measurement, \name is the first app that provides per-app network performance on unrooted phones without user intervention.
With \name, we also provide the first report of large-scale per-app network measurements.


Recently, researchers are interested in utilizing the \texttt{VpnService} API for different purposes.
Nearly all of them focus on detecting privacy leakage~\cite{TaintDroid10} by relaying and intercepting mobile apps' traffic either in the smartphone~\cite{PrivacyGuard15, Haystack15} or at a remote VPN server~\cite{AntMonitor15, ReCon16}.
Two recent works~\cite{DifferentiationDetector15, TrafficGuard16} use a remote VPN server to identify traffic differentiation and optimize traffic volume in cellular networks.
\name is different from all these related works in that we leverage the \texttt{VpnService} API for per-app network performance measurement.
Indeed, \name is the first and the only one on the market that provides per-app measurement for end users.
Moreover, our solutions for tackling the delayed VPN read problem
(\mysec\ref{sec:tuninput}) and mitigating the VPN \texttt{protect()}
delay (\mysec\ref{sec:socketwrite}) can benefit all VPN-based apps, such
as OpenVPN~\cite{openvpnHome}.

Due to the traffic-interception capability of \texttt{Vpn-} \texttt{Service} APIs, it is important for VPN-based apps to preserve users' privacy in their design.
Unfortunately, many VPN apps on the market fail to do so according to a recent study~\cite{VPNRisk16}.
The majority of them use remote VPN servers for traffic relay, but not always in a secure fashion (e.g., no encryption for the tunnel to VPN servers, or no tunneling for DNS traffic).
In contrast, our \name adopts the local phone-side traffic forwarding scheme, without additional risks associated with VPN servers, such as leaking user traffic.
Further, unlike PrivacyGuard~\cite{PrivacyGuard15} and Haystack that perform traffic \textit{content} inspection, \name makes no such attempt, let alone the TLS interception performed by those two.
This may be an important factor contributing to a much higher number of \name installs than Haystack, which reached only 1.5K installs by the end of March in 2017~\cite{LumenTweet1703}.

\vspace{-3ex}
\section{Conclusion}
\label{sec:conclude}
\vspace{-3ex}

In this paper we proposed \name, a novel measurement app to monitor per-app network performance on unrooted smartphones.
By leveraging the \texttt{VpnService} API on Android to intercept all network traffic, \name was able to opportunistically measure each app for its network RTT without network overhead and user intervention.
We overcame a number of challenges to achieve a fast packet relaying and an accurate measurement in \name.
We have deployed \name to Google Play for an IRB-approved crowdsourcing study for over ten months.
By collecting and analyzing the first large-scale per-app measurement dataset, we discovered a number of new findings on the per-app and DNS network performance experienced by real users in the wild.
We plan to further improve \name (e.g., supporting more metrics beyond RTT), and release more analysis results for app developers and ISPs to optimize their performance.



{\small \bibliographystyle{acm}
\bibliography{main}}

\begin{thebibliography}{10}

\bibitem{SocketRegister}
{AbstractSelectableChannel selector() | Android Developers}.
\newblock
  \url{http://developer.android.com/reference/java/nio/channels/spi/AbstractSelectableChannel.html#register(java.nio.channels.Selector,
  int, java.lang.Object)}.

\bibitem{dashboards}
{Dashboards | Android Developers}.
\newblock \url{https://developer.android.com/about/dashboards/index.html}.

\bibitem{DownloadManager}
{DownloadManager | Android Developers}.
\newblock
  \url{https://developer.android.com/reference/android/app/DownloadManager.html}.

\bibitem{LumenTweet1703}
{Lumen Privacy Monitor has reached 1.5K installs on Google Play!}
\newblock \url{https://twitter.com/lumen_app/status/845230899226689537}.

\bibitem{mobiperf_android}
{MobiPerf on Google Play}.
\newblock \url{https://play.google.com/store/apps/details?id=com.mobiperf}.

\bibitem{MopEyeOnPlay}
{MopEye on Google Play}.
\newblock \url{https://play.google.com/store/apps/details?id=com.mopeye}.

\bibitem{netalyzr_android}
{Netalyzr on Google Play.}
\newblock
  \url{https://play.google.com/store/apps/details?id=edu.berkeley.icsi.netalyzr.android}.

\bibitem{nspeedtest_win}
{Network Speed Test on Windows Store.}
\newblock
  \url{http://www.windowsphone.com/en-us/store/app/network-speed-test/9b9ae06b-2961-41ef-987d-b09567cffe70}.

\bibitem{openvpnHome}
{OpenVPN - Open Source VPN}.
\newblock \url{https://openvpn.net/}.

\bibitem{RFC1323}
{RFC 1323 - TCP Extensions for High Performance}.
\newblock \url{http://tools.ietf.org/html/rfc1323}.

\bibitem{RFC793}
{RFC 793 - Transmission Control Protocol}.
\newblock \url{https://tools.ietf.org/html/rfc793}.

\bibitem{Singtel4GHistory}
{Singtel's 4G Network Deployment History}.
\newblock \url{https://www.singtel.com/personal/i/4g/why-singtel}.

\bibitem{speedchecker_android}
{SpeedChecker on Google Play.}
\newblock
  \url{https://play.google.com/store/apps/details?id=uk.co.broadbandspeedchecker}.

\bibitem{speedtestx_ios}
{Speedtest X HD WiFi \& Mobile Speed Test on App Store.}
\newblock
  \url{https://itunes.apple.com/us/app/speedtest-x-hd-wifi-mobile/id366593092}.

\bibitem{speedtest_ios}
{Speedtest.net on App Store.}
\newblock
  \url{https://itunes.apple.com/us/app/speedtest.net-mobile-speed/id300704847}.

\bibitem{speedtest_android}
{Speedtest.net on Google Play.}
\newblock
  \url{https://play.google.com/store/apps/details?id=org.zwanoo.android.speedtest}.

\bibitem{speedtest_win}
{Speedtest.net on Windows Store.}
\newblock
  \url{http://www.windowsphone.com/en-us/store/app/speedtest-net/4fcd4de1-050b-44dc-b123-a786808eb49b}.

\bibitem{VPNprotect}
{The protect() API in VpnService | Android Developers}.
\newblock
  \url{http://developer.android.com/reference/android/net/VpnService.html#protect(int)}.

\bibitem{ToyVpn}
{ToyVpn}.
\newblock
  \url{https://github.com/android/platform_development/tree/master/samples/ToyVpn}.

\bibitem{AndroidVPNAPI}
{VpnService | Android Developers}.
\newblock
  \url{http://developer.android.com/reference/android/net/VpnService.html}.

\bibitem{Latest4G}
{What are 4G, 4G+ and Tri-band 4G+?}
\newblock \url{https://www.singtel.com/personal/i/4g/support}.

\bibitem{IMC14}
{\sc Baltrunas, D., Elmokashfi, A., and Kvalbein, A.}
\newblock Measuring the reliability of mobile broadband networks.
\newblock In {\em Proc. ACM IMC\/} (2014).

\bibitem{PktLoss15}
{\sc Baltrunas, D., Elmokashfi, A., and Kvalbein, A.}
\newblock Dissecting packet loss in mobile broadband networks from the edge.
\newblock In {\em Proc. IEEE INFOCOM\/} (2015).

\bibitem{QoEDoctor14}
{\sc Chen, Q., Luo, H., Rosen, S., Mao, Z., Iyer, K., Hui, J., Sontineni, K.,
  and Lau, K.}
\newblock {QoE Doctor}: Diagnosing mobile app {QoE} with automated {UI} control
  and cross-layer analysis.
\newblock In {\em Proc. ACM IMC\/} (2014).

\bibitem{WiFiLTE14}
{\sc Deng, S., Netravali, R., Sivaraman, A., and Balakrishnan, H.}
\newblock {WiFi}, {LTE}, or both?: Measuring multi-homed wireless {Internet}
  performance.
\newblock In {\em Proc. ACM IMC\/} (2014).

\bibitem{TaintDroid10}
{\sc Enck, W., Gilbert, P., Chun, B., Cox, L., Jung, J., McDaniel, P., and
  Sheth, A.}
\newblock {TaintDroid}: An information-flow tracking system for realtime
  privacy monitoring on smartphones.
\newblock In {\em Proc. Usenix OSDI\/} (2010).

\bibitem{First10}
{\sc Falaki, H., Lymberopoulos, D., Mahajan, R., Kandula, S., and Estrin, D.}
\newblock A first look at traffic on smartphones.
\newblock In {\em Proc. ACM IMC\/} (2010).

\bibitem{4GTest12}
{\sc Huang, J., Qian, F., Gerber, A., Mao, Z., Sen, S., and Spatscheck, O.}
\newblock A close examination of performance and power characteristics of {4G}
  {LTE} networks.
\newblock In {\em Proc. ACM MobiSys\/} (2012).

\bibitem{LTE13}
{\sc Huang, J., Qian, F., Guo, Y., Zhou, Y., Xu, Q., Mao, Z., Sen, S., and
  Spatscheck, O.}
\newblock An in-depth study of {LTE}: Effect of network protocol and
  application behavior on performance.
\newblock In {\em Proc. ACM SIGCOMM\/} (2013).

\bibitem{3GTest10}
{\sc Huang, J., Xu, Q., Tiwana, B., Mao, Z., Zhang, M., and Bahl, P.}
\newblock Anatomizing application performance differences on smartphones.
\newblock In {\em Proc. ACM MobiSys\/} (2010).

\bibitem{VPNRisk16}
{\sc Ikram, M., Vallina-Rodriguez, N., Seneviratne, S., Kaafar, M.~A., and
  Paxson, V.}
\newblock An analysis of the privacy and security risks of {Android VPN}
  permission-enabled apps.
\newblock In {\em Proc. ACM IMC\/} (2016).

\bibitem{JavaSelector}
{\sc Jenkov, J.}
\newblock {Java NIO Selector}.
\newblock \url{http://tutorials.jenkov.com/java-nio/selectors.html}.

\bibitem{Bufferbloat12}
{\sc Jiang, H., Liu, Z., Wang, Y., Lee, K., and Rhee, I.}
\newblock Understanding bufferbloat in cellular networks.
\newblock In {\em Proc. ACM CellNet\/} (2011).

\bibitem{DifferentiationDetector15}
{\sc Kakhki, A., Razaghpanah, A., Li, A., Koo, H., Golani, R., Choffnes, D.,
  Gill, P., and Mislove, A.}
\newblock Identifying traffic differentiation in mobile networks.
\newblock In {\em Proc. ACM IMC\/} (2015).

\bibitem{NornetEdge14}
{\sc Kvalbein, A., Baltrunas, D., Evensen, K., Xiang, J., Elmokashfi, A., and
  Oliveira, S.}
\newblock The {Nornet Edge} platform for mobile broadband measurements.
\newblock In {\em Computer Networks\/} (2014).

\bibitem{AntMonitor15}
{\sc Le, A., Varmarken, J., Langhoff, S., Shuba, A., Gjoka, M., and
  Markopoulou, A.}
\newblock {AntMonitor}: A system for monitoring from mobile devices.
\newblock In {\em ACM SIGCOMM Workshop on Crowdsourcing and Crowdsharing of Big
  Internet Data (C2BID)\/} (2015).

\bibitem{wchli15}
{\sc Li, W., Mok, R., Wu, D., and Chang, R.}
\newblock On the accuracy of smartphone-based mobile network measurement.
\newblock In {\em Proc. IEEE INFOCOM\/} (2015).

\bibitem{CoNEXT16}
{\sc Li, W., Wu, D., Chang, R., and Mok, R. K.~P.}
\newblock Demystifying and puncturing the inflated delay in smartphone-based
  {WiFi} network measurement.
\newblock In {\em Proc. ACM CoNEXT\/} (2016).

\bibitem{TrafficGuard16}
{\sc Li, Z., Wang, W., Xu, T., Zhong, X., Li, X.-Y., Liu, Y., Wilson, C., and
  Zhao, B.~Y.}
\newblock Exploring cross-application cellular traffic optimization with {Baidu
  TrafficGuard}.
\newblock In {\em Proc. USENIX NSDI\/} (2016).

\bibitem{Mobilyzer15}
{\sc Nikravesh, A., Yao, H., Xu, S., Choffnes, D., and Mao, Z.}
\newblock Mobilyzer: An open platform for controllable mobile network
  measurements.
\newblock In {\em Proc. ACM MobiSys\/} (2015).

\bibitem{RRC3G10}
{\sc Qian, F., Wang, Z., Gerber, A., Mao, Z., Sen, S., and Spatscheck, O.}
\newblock Characterizing radio resource allocation for {3G} networks.
\newblock In {\em Proc. ACM IMC\/} (2010).

\bibitem{Haystack15}
{\sc Razaghpanah, A., Vallina{-}Rodriguez, N., Sundaresan, S., Kreibich, C.,
  Gill, P., Allman, M., and Paxson, V.}
\newblock Haystack: In situ mobile traffic analysis in user space.
\newblock {\em CoRR abs/1510.01419\/} (2015).

\bibitem{ReCon16}
{\sc Ren, J., Rao, A., Lindorfer, M., Legout, A., and Choffnes, D.~R.}
\newblock {ReCon}: Revealing and controlling {PII} leaks in mobile network
  traffic.
\newblock In {\em Proc. ACM MobiSys\/} (2016).

\bibitem{RRCState14}
{\sc Rosen, S., Luo, H., Chen, Q., Mao, Z., Hui, J., Drake, A., and Lau, K.}
\newblock Discovering fine-grained {RRC} state dynamics and performance impacts
  in cellular networks.
\newblock In {\em Proc. ACM MobiCom\/} (2014).

\bibitem{CellDNS14}
{\sc Rula, J.~P., and Bustamante, F.~E.}
\newblock Behind the curtain: Cellular {DNS} and content replica selection.
\newblock In {\em Proc. ACM IMC\/} (2014).

\bibitem{PrivacyGuard15}
{\sc Song, Y., and Hengartner, U.}
\newblock {PrivacyGuard}: A {VPN}-based platform to detect information leakage
  on {Android} devices.
\newblock In {\em ACM CCS Workshop on Security and Privacy in Smartphones and
  Mobile Devices (SPSM)\/} (2015).

\bibitem{ProfileDroid12}
{\sc Wei, X., Gomez, L., Neamtiu, I., and Faloutsos, M.}
\newblock {ProfileDroid}: multi-layer profiling of {Android} applications.
\newblock In {\em Proc. ACM MobiCom\/} (2012).

\bibitem{MopEyePoster15}
{\sc Wu, D., Li, W., Chang, R., and Gao, D.}
\newblock {MopEye}: Monitoring per-app network performance with zero
  measurement traffic.
\newblock In {\em Proc. ACM CoNEXT Student Workshop\/} (2015).

\bibitem{IdentifyBehavior11}
{\sc Xu, Q., Erman, J., Gerber, A., Mao, Z., Pang, J., and Venkataraman, S.}
\newblock Identifying diverse usage behaviors of smartphone apps.
\newblock In {\em Proc. ACM IMC\/} (2011).

\end{thebibliography}

\end{document}